\documentclass[11pt]{JHEP3}
\usepackage{graphicx}
\usepackage{amssymb}

\def\ba{\begin{eqnarray}}
\def\ea{\end{eqnarray}}
\def\be{\begin{equation}}
\def\ee{\end{equation}}
\def\ben{\begin{equation} \nonumber}
\def\een{\end{equation}}
\def\baray{\begin{eqnarray*}}
\def\earay{\end{eqnarray*}}

\def\half{{1\over2}}

\def\phs{\phi_S}
\def\phl{\phi_L}
\def\ps{\pi_S}
\def\pl{\pi_L}
\def\etap{\eta_{\phi}}

\def\phil{\frac{d\phi_{\alpha}}{d \lambda}}
\def\pil{\frac{d\pi_{\alpha}}{d \lambda}}

\def\dHdpi{\frac{\partial H_{\rm eq}}{\partial \pi_{\alpha}}}
\def\pii{\pi_{\alpha}}
\def\phii{\phi_{\alpha}}
\def\pij{\pi_{\beta}}
\def\phij{\phi_{\beta}}

\def\dl{d \lambda}

\def\({\left(}
\def\){\right)}

\title{Stochastic Inflation Revisited: \\ Non-Slow Roll Statistics and DBI Inflation}
\author{Andrew J. Tolley${}^{\rm a}$ and Mark Wyman${}^{\rm b}$\\
	 Perimeter Institute for Theoretical Physics \\
	 31 Caroline St. N \\
	 Waterloo, ON  N2L 2Y5 \\
	 Canada\\
	 ${}^{\rm a}$ \email{atolley@perimeterinstitute.ca} \, ${}^{\rm b}$ \email{mwyman@perimeterinstitute.ca}}

\abstract{Stochastic inflation describes the global structure of the
inflationary universe by modeling the super-Hubble dynamics as a
system of matter fields coupled to gravity where the
sub-Hubble field fluctuations induce a stochastic force into the equations of motion.
The super-Hubble dynamics are ultralocal, allowing us to neglect
spatial derivatives and treat each Hubble patch as a separate universe.
This provides a natural framework in which to discuss probabilities
 on the space of solutions and initial conditions. In this
article we derive an evolution equation for this probability for an arbitrary class of matter systems, including
DBI and k-inflationary models, and discover equilibrium solutions that satisfy detailed balance.
Our results are more general than those derived assuming slow roll
or a quasi-de Sitter geometry, and so are directly applicable to models that do not satisfy the usual slow roll conditions. We discuss in general terms the conditions for eternal inflation to set in, and we give explicit numerical solutions of highly stochastic, quasi-stationary trajectories in the relativistic DBI regime.
Finally, we show that the probability for stochastic/thermal tunneling can be significantly enhanced relative to the Hawking-Moss instanton result due to relativistic DBI effects. }

\preprint{PI-COSMO-58}
\begin{document}

\section{Introduction}

The great success of the inflationary paradigm has given impetus to
each of the areas of physics upon which it touches. Its essential character,
though, is captured by the dynamics of a scalar degree of freedom that
sources gravity in such a way as to generate an accelerating expansion
rate for the Universe. The basic predictions
of inflation require only an understanding of the classical dynamics
of a scalar field in the overdamped limit. Though simple, meeting this requirement
has launched a thousand models and remains an active area of research.
Theoretical discrimination among models is difficult to achieve, particularly when some
models possess exotic features like non-standard kinetic terms. Even within models, it is difficult to say with certainty whether particular inflationary trajectories are natural or contrived.

It is no easy task to find an approach general enough to encompass the wide
variety of inflationary models that still provides useful insight into
inflationary dynamics. One way forward comes from taking a statistical approach,
a technique that goes under the name of stochastic inflation.
Pioneered by Starobinsky \cite{starobinsky},
stochastic inflation treats sub-Hubble scalar field fluctuations as a stochastic source of noise for the mean value of the field coarse-grained over a Hubble volume. The noise crosses the horizon
during the inflationary expansion, allowing it to influence the super-Hubble behavior
of the field. Studying this interplay allows us to determine the evolution of probability density
for particular solutions to the field equations, a necessary first step towards
understanding which inflationary trajectories are likely and which are not.

There has been a great deal of work done applying Starobinsky's methods
to a variety of inflationary scenarios and in understanding other stochastic effects for
inflationary fields \cite{Goncharov:1987ir, Rey:1987,Sasaki:1988a,Hosoya:1988b,Graziani,Linde:1993xx, Prokopec:2007ak}. Readers interested in a review of stochastic
techniques and applications are referred to \cite{Linde:2005ht}. Some recent
progress includes finding an analog of Starobinsky's
celebrated Fokker-Planck equation for Dirac-Born-Infeld (DBI) inflation \cite{Silverstein:2003hf}
in Ref.~\cite{Chen:2006hs}; also, probabilities and stationary solutions in k-inflation \cite{ArmendarizPicon:1999rj}
were studied in Ref.~\cite{Helmer:2006tz}.
In this work, we describe general techniques that capture the stochastic behavior of inflating
fields without relying on slow roll.  This allows us to
treat either exotic DBI or k-essence fields and canonical fields
within a unified framework.

To make clear what reductions of the full dynamics we make, we shall initially formulate the stochastic system on the full phase space. There are two ways in which the coarse-graining between super-Hubble and sub-Hubble modes can be achieved, which we refer to as the one- and two-noise approaches. In the one-noise approach, the stochastic noise is viewed as an additional force in the classical equations of motion, so that the momentum receives kicks, but the time derivative of the field is related to the momentum according to its classical equation. In the two-noise approach both the field and momentum receive stochastic kicks. The two approaches are qualitatively (although not identically) the same. The two-noise approach has the advantage that if the sub-Hubble perturbations are assumed to be adiabatic, then it is equivalent  to the stochastic Hamilton-Jacobi approach considered in Ref.~\cite{Chen:2006hs} after a phase space transformation. A disadvantage to this approach is that when applied to models such as DBI inflation in which the scalar has a speed limit, stochastic jumps will typically violate this limit, even though on average it is satisfied. By contrast, the speed limit continues to be satisfied even with stochastic fluctuations included in the one-noise approach.

In the two-noise case we can find a one parameter family of equilibrium solutions which satisfy detailed balance. This additional parameter is a reflection of the fact that the solutions retain a partial memory of their initial state. This is the same parameter that arises as the integration constant in solving the Hamilton-Jacobi equation, so it is fixed by the initial conditions for the field and momentum.
One of the most important reasons to understand whether equilibrium solutions exist
in any model of inflation is to understand stochastic tunneling, i.e. tunneling described by the Hawking-Moss instanton \cite{Hawking:1981fz}. If the inflationary potential has a variety of metastable
minima for which the tunneling rate to other vacua is low, there is usually sufficient time for a partial equilibrium to set in before the field tunnels to the true vacuum.
The equilibrium solutions we find through our stochastic approach
allow us to derive the transition rates for thermally-activated/stochastic tunneling, thus generalizing the Hawking-Moss instanton result without making use of difficult to interpret Euclidean gravity techniques.
Qualitatively, what we find is a transition amplitude similar to the usual one when $c_s$, the effective sound speed of the perturbations, is everywhere close to unity;
however, in regions of phase space for which $c_s \ll 1$, we find a significantly enhanced tunneling rate. This is of significant interest in understanding tunneling in models of the so-called landscape of string theory (cf. \cite{HenryTye:2006tg,Podolsky:2007vg,Sarangi:2007jb,Copeland:2007qf}). Furthermore, the tunneling rate, like the equilibrium solution, retains a partial memory of the initial probability state.

One practical way to make use of our framework is to use it to give
explicit solutions for stochastic inflationary trajectories. We did this for DBI inflation.
In agreement with earlier results in \cite{Chen:2006hs}, we find that in a phenomenologically viable DBI model with only an $m^2 \phi^2$ potential, eternal inflation is not possible, at least in the relativistic DBI regime. However, by adding higher order polynomial corrections to the potential,
which are expected to kick in at large field values,
we can find solutions that possess  both a phenomenologically viable regime at small values of $\phi$ and a stochastic DBI regime at large values of $\phi$. Intriguingly, for certain initial conditions the evolution of $\phi$ in this stochastic regime deviates strongly from its classical solution. 
The field undergoes an apparently metastable random walk, yet with $c_s$ on average well below unity. This situation is quite different from the stochastic regimes encountered in minimally coupled models in that the relativistic ``speed limit", rather than
Hubble damping, is setting the field's dynamics. Since brane inflation and other uses of the DBI
action are still under active development, it will be important to determine whether regimes
that exhibit these dynamics exist in fully worked out constructions. If they do, it will be
necessary for stochastic dynamics to be taken into account when determining what inflationary
trajectories are available and likely.

Although our generalized stochastic approach does not provide any unambiguous
measure for inflation, it does allow us to state clearly the assumptions behind
recent proposals for a measure. For instance, the phase space measure proposed by Gibbons
and Turok \cite{GT} assumes that stochastic diffusion is absent from the scalar field
evolution, and so including the stochastic noise tells us precisely how this measure will be adjusted when fluctuations are included. We find that applying their classical arguments to DBI inflation gives precisely the same $e^{-3N}$ suppression. Alternatively, one can incorporate volume weighting into the Fokker-Planck equation, although this procedure is not manifestly gauge invariant.  Recent proposals have attempted to improve on this (see e.g. \cite{Hartle:2007gi,Linde:2007nm}).

We shall give a somewhat pedagogical discussion, emphasizing a number of standard results not often found together in the literature. For instance, we demonstrate that the stochastic approach requires neither a nearly de Sitter geometry nor a slowly rolling field, but only that the modes are exiting the horizon and the growing mode is dominant, and can be formulated in way that consistently incorporates gravitational backreaction. We begin in Sec.~\ref{PSFP} with derivation methods for both the Langevin and Fokker-Planck equations. We emphasize the crucial role of the consistent ultralocal truncation of GR coupled to matter fields with its resulting scaling symmetry -- which provides a natural definition of time through e-folds of volume increase. We apply  these methods to DBI inflation in Sec.~\ref{DBIinflation}. We then discuss several of the properties and consequences of the stochastic dynamics in Sec.~\ref{properties}, including a general prescription for when quantum corrections come to dominate over classical ones -- a necessary prerequisite for eternal inflation. In Sec.~\ref{numerics} we numerically calculate some explicit stochastic trajectories for DBI inflation.
 We focus on trajectories exhibiting
strongly stochastic behaviour in a highly relativistic regime. Most intriguing is our discovery that
apparently metastable ultra-relativistic random walks exist for super-Planckian field values.
For these fields, the value of $\phi$ does not systematically roll in any particular direction,
yet the field's mean sound speed, $c_s$, is well below unity. Finally, in Sec. \ref{tunneling}, we calculate thermally-activated
tunneling rates using our stochastic approach and demonstrate the above-mentioned
enhancement in tunneling rate
due to $c_s < 1$ effects. In Appendix A
we demonstrate that the Hubble parameter is the generator of e-folding time-translations for ultralocal gravity and
in Appendix B we give a formal path integral translation between the Langevin (``equation of motion") and Fokker-Planck
(``conservation of probability density") approaches to stochastic dynamics.

\section{Phase Space Fokker-Planck Equation}

\label{PSFP}

\subsection{The importance of ultralocality}
\label{scaling}
A central ingredient to the stochastic approach is to consider the ultralocal limit
of the action for general relativity coupled to matter fields. This limit is a consistent truncation of GR that arises when time derivatives dominate over spatial derivatives.
This reduced phase space exhibits an important new feature:  a
scaling symmetry. This can be viewed as a time translation invariance in e-folding time. It is this symmetry
that allows the conservation of curvature perturbations on super-horizon scales.
Time translation invariance also permits us to develop a concept of
detailed balance for the probability equations we subsequently derive.

Following Salopek and Bond ~\cite{Salopek:1990a, Salopek:1991a},
we describe the line element for our space-time in the Arnowitt-Deser-Misner form,
\be
ds^2 =- N^2 dt^2 + \gamma_{ij} (dx^i + N^i dt)(dx^j + N^j dt)
%g_{00}= -N^2+ \gamma^{ij}N_iN_j,\; g_{0i} = g_{i0} = N^i, \; g_{ij}=\gamma_{ij},
\ee
where $N$ and $N_i$ are the lapse and shift functions, and $\gamma_{ij}$
is a general three-metric.
We consider the action for gravity coupled to a collection of non-minimal kinetic term scalar matter fields $\phi_a$ (used here as a stand
in for general matter fields) given by,
\be
S  =  \int d^4xN \sqrt{\gamma} \left \{ \frac{M_{P}^2}{2} \left[ ^{(3)}R+K_{ij}K^{ij} - K^2\right] +    p(X,\phi_a) \right \} ,
\ee
where $p(X,\phi_a)$ is a general scalar field Lagrangian and
$$
X=\frac{1}{2}G^{ab} \left[ N^{-2}  (\dot{\phi}_a - N^i\phi_{a;i}) (\dot{\phi}_b, - N^i\phi_{b;i}) -\phi_{a;i}\phi_b^{;i}\right],
$$
where $^{(3)}R$ is the three-space curvature associated with $\gamma_{ij}$, $G_{ab}$ is the metric
on field space (a Kronecker delta for canonical scalar fields), $M_{P}^2$ is the reduced Planck mass, and
\be
K_{ij} = {1\over{2N}}\left ( -N_{i;j} - N_{j;i} + { {\partial \gamma_{ij}}\over{\partial t}} \right ), \quad K = K^i_i.
\ee
It is useful to define scalar and gravitational momenta,
\be \label{momenta}
\pi^a = G^{ab}N^{-1}(\dot{\phi}_a - N^{i}\phi_{a,i}) \, p_{,X}, \quad [\pi^\gamma]^{ij} = (M_{P}^2/2)(K^{ij}-\gamma^{ij} K ),
\ee
where, for future convenience, these expressions differ from the standard definitions ($p_{\alpha}$) (e.g.  \cite{Salopek:1991gr}) by a volume factor $\pi_{\alpha}=p_{\alpha}/\sqrt{\gamma}$. In  terms of these momenta, we can recast the action as
\be \label{saloaction}
S = \int d^4x \sqrt{\gamma} \; \(g^{ab} \pi^{\phi_a} \dot{\phi}_b + [\pi^\gamma]^{ij} \dot{\gamma}_{ij} - N \mathcal{H} - N^{i} \mathcal{H}_i\),
\ee
where the energy density $\mathcal{H}$ and a momentum
density $\mathcal{H}_i$ are:
\baray
0  & = \mathcal{H} & =   2 M_{P}^{-2} [ \gamma_{jk} \gamma_{li} [\pi^\gamma]^{ij}
[\pi^\gamma]^{kl} - \half (\pi^{\gamma})^2] +\frac{G_{ab}\pi^a\pi^b}{p_{,X}(X_0,\phi_a)}-p(X_0,\phi_a)\\
&  + &  \left (- \frac{M_{P}^{2}}{2} \;^{(3)}R +  \frac{G_{ab}\pi^a\pi^b}{p_{,X}(X,\phi_a)}-p(X,\phi_a)-\frac{G_{ab}\pi^a\pi^b}{p_{,X}(X_0,\phi_a)}+p(X_0,\phi_a)\right )
	 \leftarrow \mbox{neglect as small}\\
0 & = \mathcal{H}_i & =   - 2(\gamma_{il}[\pi^\gamma]{}^{lk}),_k + [\pi^\gamma]^{lk} \gamma_{lk}{}_{;i} + \pi^{\phi_k} \phi_k{}_{;i},
\earay
and we have written $\pi^\gamma \equiv [\pi^\gamma]^{ij} \gamma_{ij} = Tr[\pi^\gamma]^{i}_j$. Here $X$ and $X_0$ are implicitly defined by the relations
\be
X = \frac{1}{2}G_{ab}\frac{\pi^a\pi^b}{p_{,X}^2(X,\phi_a)}-\frac{1}{2}G^{ab}\phi_{a;i}\phi_b^{;i}, \quad
X_0 = \frac{1}{2}G_{ab}\frac{\pi^a\pi^b}{p_{,X}^2(X_0,\phi_a)}.
\ee
We define $X$ -- the total kinetic term -- and $X_0$ -- the temporal part of the kinetic term --
so that the ultralocal limit is consistently applied to the nonstandard kinetic terms encompassed
by the $p(X,\phi)$ formalism. 
The neglect of the terms in the large parenthesis is precisely the statement of ultralocality: the limit
in which each point in space evolves independently of the points around it, as if it were its own separate universe.
This is achieved here by neglecting all terms in the Hamiltonian which are second order or higher
in spatial gradients.
Crucially, this limit is not the same as the assumption of isotropy and local homogeneity:
 the momentum constraint is untouched by this reduction
of the problem and contains crucial information on how neighboring Hubble-patches are pieced together.
In the special case of minimally coupled fields, Salopek \cite{Salopek:1991gr} demonstrated that the reduced constraints ${\cal H}$ and ${\cal H}_i$ still form a consistent constraint algebra, and thus represent a kind of contraction of the full diffeomorphism group that nevertheless includes the same number of generators. It is not hard to see that these statements generalize to gravity coupled to any matter system that respects general covariance, and in particular to the non-minimal kinetic models that we shall consider in this paper.
Though we do not pursue this avenue at present, we wish to point out
that the ultralocal limit retains superhorizon tensor mode dynamics.
Since gravity waves, which are associated with this freedom,
are not dependent on any potential, they simply give an additive contribution to the overall Hamiltonian, as seen above, and their dynamics are somewhat trivial even when stochastic fluctuations are included. For this reason they are usually neglected.

\subsubsection{Scaling Symmetry in the ultralocal Limit}

Once the terms in the large parenthesis have been neglected, we have
a system that is invariant under the arbitrary scaling
\begin{eqnarray}
\gamma_{ij} & \rightarrow  e^{2c} \gamma_{ij} & \quad \pi_a \rightarrow  \pi_a \nonumber \\
N & \rightarrow  N &  \quad [\pi^\gamma]^i_j \rightarrow [\pi^\gamma]^i_j \nonumber \\
N^{i} & \rightarrow  N^i & \quad \mbox{thus, }S \rightarrow e^{3c} S. \label{scalingeq}
\end{eqnarray}
The crucial consequence of this symmetry is that the equations of motion are invariant under these transformations.
In what follows, we will write $\sqrt{\gamma} \rightarrow e^{3\lambda}$, where $\lambda$ is
the number of e-folds of expansion (counting forwards in time in an expanding universe). We shall find it convenient to use the number of e-folds as our time-coordinate. With this choice of time, the scaling symmetry mentioned
here becomes, explicitly, a {\it time translation invariance}:
the equations of motion for $\lambda + c$ are equivalent to those at $\lambda$. Here we reap the first benefit of making the assumption of ultralocality. Given a field $\phi(\lambda)$ that satisfies the equations of
motion, then it automatically follows that $\phi(\lambda+c)$ satisfies the equations of motion. We will then be guaranteed
that one perturbative mode will satisfy
\be
\delta \phi_a = \phi_a(\lambda + c) - \phi_a(\lambda) \propto \frac{d\phi_{a}}{d\lambda}.
\ee
When we construct the (physical observable) curvature perturbation, $\zeta$, arising from this mode, we find
that it is conserved:
\be
\label{linearcons}
\zeta = \psi + H \frac{\delta\phi_a}{\frac{d \phi_a}{dt}} = \frac{\delta \phi_a}{\frac{d \phi_{a}}{d\lambda}} \simeq \mbox{conserved!}
\ee
where $\psi =0$ in our choice of gauge. The conservation
of $\zeta$ is usually attributed to energy conservation for modes outside of the horizon~\cite{Constancy}; here, however,
we can see that this conservation reflects the time translation invariance of super-Hubble modes.
We can extend this linear argument to include non-linear effects in  a straightforward way, and do
so in \S \ref{Hamilton-Jacobi}.
The constancy of one mode of $\zeta$ at linear order is generic; that is, there is always a curvature perturbation mode that is
conserved outside the horizon. However, the existence of one constant mode does not mean that $\zeta$ is always conserved since entropy modes can mix into the constant part of $\zeta$, and even temporarily dominate over the constant mode. This occurs, for example, for curvaton models~\cite{curvaton}, or for multifield systems in collapsing spacetimes \cite{Tolley:2007nq}.

In what follows, we will consistently solve the temporal constraint equation ${\mathcal H}=0$, thus removing
it from consideration. In practice we achieve this by solving ${\mathcal H}=0$ for $\pi_{\lambda}=Tr(\pi^{\gamma})$, and substituting back into the action.
Combined with fixing the time-reparameterization invariance by choosing e-folding time, we are left with only the spatial constraint ${\cal H}_i$ and the associated spatial diffeomorphism invariance.
However, ultralocality is precisely the statement that we do not consider spatial variations in the dynamical equations. Consequently, this constraint plays no role in the long wavelength dynamics, but only tells us how to stitch neighboring Hubble patches together consistently.

From now on we shall work in e-folding time, in which it is straightforward to show that the dynamics of a set of fields $\phi_{\alpha}$ (not necessarily scalars) follows from an
action of the form
$$
S = \int d\lambda \int d^3x \, e^{3\lambda} \(\pi_{\alpha} \frac{\partial{\phi_{\alpha}}}{\partial \lambda} -H_{\rm eq}(\phi_{\alpha},\pi_{\alpha})-N^i {\cal H}_i\),
$$
where the volume factor $e^{3\lambda}$ has its origin in $\sqrt{-g}$, and the momentum conjugates are defined, as before, without any dependence on this volume factor. To reiterate, this is the same
action as in Eqn.~(\ref{saloaction}), rewritten using e-folding time, where we have explicitly solved for the constraint ${\cal H}=0$. For brevity we have folded the tensor degrees of freedom into our single field variable, $\phi_{\alpha} = \{\phi_a, \gamma^{k}_{l}\}$.  As we showed in Eqs. (\ref{scalingeq}), this action scales as $S \rightarrow e^{3c} S$ under the
 shift symmetry, and so the equations of motion are correspondingly invariant.

The equations of motion, given by $$ \frac{\partial{\phi_{\alpha}}}{\partial \lambda}=\frac{\partial H_{\rm eq}}{\partial \pi_{\alpha}}, \quad
\frac{\partial {\pi_{\alpha}}}{\partial \lambda}=-\frac{\partial H_{\rm eq}}{\partial \phi_{\alpha}} - 3 \pi_{\alpha}, \quad {\cal H}_i=0 ,$$ take the form of a damped Hamiltonian system,
and thus naturally split into a time reversible part  -- which follows from the undamped time translation invariant action,
$$
S_{\rm rev} = \int d\lambda \int d^3x\(\pi_{\alpha} \frac{\partial {\phi_{\alpha}}}{\partial \lambda} -H_{\rm eq}(\phi_{\alpha},\pi_{\alpha})-N^i {\cal H}_i \),
$$
which by itself would conserve $H_{\rm eq}$ -- and an irreversible part from the damping. The time reversal symmetry acts as
$\lambda \rightarrow -\lambda$, $\phi_{\alpha} \rightarrow \epsilon_{\alpha} \phi_{\alpha}$, $\pi_{\alpha} \rightarrow - \epsilon_{\alpha} \pi_{\alpha}$ (where
$\epsilon_{\alpha} = \pm 1$ is the intrinsic time parity of the field $\phi_{\alpha}$) and the Hamiltonian transforms as
$H_{\rm eq}(\phi_{\alpha},\pi_{\alpha})=H_{\rm eq}(\phi_{\alpha},-\pi_{\alpha})$. The loss of `energy' through damping is given explicitly by \be
\frac{\partial H_{\rm eq}}{\partial \lambda}=-3\pi_{\alpha}\frac{\partial H_{\rm eq}}{\partial \pi_{\alpha}} .\ee

To illustrate this, let us drop the spatial dependence and consider the canonical example of a minimally coupled scalar field and gravity whose dynamics is governed by the homogeneous and isotropic (mini-superspace) action
 \be \label{mini}
 S = \int d t e^{3  \lambda} \left \{ -\frac{3M_{P}^2\dot{\lambda}^2}{N} + \frac{\dot{\phi}^2}{2N} - N V(\phi)\right\},
 \ee
 where a dot indicates differentiation with respect to an arbitrary time coordinate, $t$. If we vary
with respect to the lapse function $N$ to find the constraint equation, namely the Friedmann equation
\be
\label{mini2}
3M_{P}^2\frac{\dot{\lambda}^2}{N^2}= \frac{\dot{\phi}^2}{2N^2} + V(\phi),
\ee
 we can solve directly for the lapse
function $N$: $$\label{lapse1}
N=\dot{\lambda}M_{P} \sqrt{\frac{3}{V(\phi)}\( 1-\frac{1}{6M_{P}^2}\left (\frac{d\phi}{d\lambda}\)^2\right )}.$$
Substituting this directly back into the action Eq.~(\ref{mini}), we find
\be
S =\int d\lambda \, {\mathcal L}= -2M_{P}  \int d\lambda \underbrace{e^{3
\lambda}}_{\rm irrev} \underbrace{ \sqrt{3V(\phi)\(1 - \frac{1}{6M_{P}^2}\(\frac{d\phi}{d\lambda}\)^2\) }}_{\rm reversible}.
\ee
This step effectively removes the residual diffeomorphism invariance from the system and reduces it to that of a non-minimally coupled damped scalar field system exhibiting the promised scaling symmetry. We have labeled the parts of the resulting action that are
time reversible and time irrev(ersible).  In this case the conjugate momentum is given by
\be \pi_{\phi}= e^{-3\lambda}\frac{\partial {\mathcal L}}{\partial (\frac{\partial \phi}{\partial
\lambda})}=\frac{V(\phi)}{M_{P} \sqrt{ 3V(\phi)\(1 - \frac{1}{6M_{P}^2}\(\frac{d\phi}{d\lambda} \)^2\) }} \frac{d\phi}{d\lambda}, \ee and so the equilibrium Hamiltonian is given by
\be
H_{\rm eq}=\pi_{\phi} \phi,_\lambda-e^{-3\lambda} {\mathcal L}=6M_{P}^2\sqrt{\frac{\pi_{\phi}^2}{6}+\frac{V(\phi)}{3}} .
\ee
One can demonstrate that the damped
Hamiltonian equations --  given
explicitly by $$\frac{d \phi}{d \lambda} =\frac{6(M_{P}^2)^2}{H_{\rm eq}} \pi_{\phi}, \quad
  \frac{d \pi_{\phi}}{d \lambda}  =-3\pi_{\phi}-\frac{6(M_{P}^2)^2}{H_{\rm eq}}V_{,\phi},$$
-- when combined with the map from e-folding time to proper time $N_{\lambda} d\lambda = 6M_{P}^2 H_{\rm eq}^{-1} d\lambda = dt$, are identical to the usual Friedmann equation and scalar field equation. It is also possible to reintroduce a spatial curvature term by making the replacement $V \rightarrow V-ke^{-2\lambda}$, although this clearly breaks the time translation invariance.

A couple of comments  about the form of this Hamiltonian are in order. Firstly, it resembles the Hamiltonian for a relativistic
particle with mass $m \propto \sqrt{V}$, which is a consequence of the fact that the mini-superspace action is equivalent to that of a particle moving on
a 1+1 dimensional manifold with space-time dependent mass. Secondly, remembering that the Hubble parameter is defined via
$H=\dot{\lambda}/N = 1/N_\lambda$ and using the equation (\ref{mini2}) for $N_{\lambda}$ we have \be H_{\rm eq}=6M_{P}^2 H. \ee This result is no coincidence, but rather a universal relationship,
true for any system in which the the action for gravity is given by the Einstein-Hilbert action, which we demonstrate in Appendix A.
Even from what we have already shown, it is clear that for a reversible system, $H_{\rm eq}$ is conserved. This implies
that the Hubble parameter, which is proportional to $H_{\rm eq}$, must also be constant. Thus, whenever a gravitational system arrives
in a state governed by reversible dynamics, it will realize a de Sitter ($H=\mbox{const}$) background. The origin of the concept of eternal inflation is precisely that sub-Hubble stochastic fluctuations can on average balance the loss of energy though damping, giving rise to precisely such an equilibrium de Sitter-like state.

\subsubsection{Relationship with Hamilton-Jacobi approach and attractor dynamics}
\label{Hamilton-Jacobi}

It is illuminating to translate our approach into the related, but more familiar, Hamilton-Jacobi formulation. They are related by a phase space (but not canonical) transformation. The trick is to reexpress the momentum, $\pi_{\alpha}$, as some function, $f_{\alpha}(\phi_{\alpha},J_{\alpha})$, of the field values $\phi_{\alpha}$ and new variables $J_{\alpha}$. The action then takes the form (neglecting spatial dependence)
\be
S = \int d\lambda \, e^{3\lambda} \(f_{\alpha}(\phi_{\beta},J_{\beta})\frac{d{\phi_{\alpha}}}{d\lambda} -\tilde{H}_{\rm eq}(\phi_{\beta},J_{\beta})\),
\ee
where $\tilde{H}_{\rm eq}(\phi_{\alpha},J_{\alpha})={H}_{\rm eq}\(\phi_{\alpha},\pi_{\alpha}=f_j(\phi_{\beta},J_{\beta})\)$. Varying with respect to $\phi_{\alpha}$ gives the equation of motion
\be
\frac{d J_{\beta}}{d \lambda} \frac{\partial f_{\alpha}}{\partial J_{\beta}}+3f_{\alpha}=-\frac{\partial \tilde{H}_{\rm eq}}{\partial \phi_{\alpha}}.
\ee
We can then simply choose $f_{\alpha}=-\frac{1}{3}\frac{\partial \tilde{H}_{\rm eq}}{\partial \phi_{\alpha}}$, which is the same as writing $\pi_{\alpha}=-2M_P^2 \frac{\partial H}{\partial \phi_{\alpha}}$ since $\tilde{H}_{\rm eq}=6M_P^2 H$.  With this choice, the variables $J_{\alpha}$ are conserved quantities
\be
\frac{d J_{\alpha}}{d \lambda}=0.
\ee
The Hamilton-Jacobi equation can then be immediately inferred from the relationship between the equilibrium Hamiltonian and the Hubble constant,
\be
H(\phi_{\alpha},J_{\alpha})=\frac{1}{6M_P^2} H_{\rm eq}\(\phi_{\alpha} , -2M_P^2  \frac{\partial H}{\partial \phi_{\alpha}}\).
\ee
The $J_{\alpha}$'s arise as integration constants in this approach,
and the Hamilton-Jacobi dynamics is encoded in the action
\be
S = \int d\lambda \, e^{3\lambda} \(-2M_P^2 \frac{\partial H}{\partial \phi_{\alpha}}\frac{d{\phi_{\alpha}}}{d\lambda} -6M_P^2 H\).
\ee

The Hamilton-Jacobi approach allows us to derive a very powerful result: It is often of interest to know whether the dynamics on phase space is an attractor. Indeed, a crucial argument in favor of inflation being insensitive to initial conditions is that it is an attractor.
Since the parameter $J_{\alpha}$ distinguishes between trajectories, let us consider the effect of a small variation in $J_{\alpha}$
\begin{eqnarray}
\delta H(\phi_{\beta},J_{\beta}) & = & \( \frac{1}{6M_P^2} \frac{\partial H_{\rm eq}}{\partial \pi_{\phi_{\alpha}}} \)\( -2M_P^2 \frac{\partial \delta H(\phi_{\beta},J_{\beta})}{\partial \phi_{\alpha}} \) \nonumber
 =  -\frac{1}{3} \frac{d\phi_{\alpha}}{d\lambda}\frac{\partial \delta H(\phi_{\beta},J_{\beta})}{\partial \phi_{\alpha}} \\ & = & -\frac{1}{3} \frac{d \delta H(\phi_{\beta},J_{\beta})}{d\lambda}.
\end{eqnarray}
That is to say, $\delta H \sim e^{-3\lambda}$ for any system -- a remarkably general result. The fact that slow roll solutions are usually attractors has its origin in this very general argument, since if the effective equation of state of the background solutions is $w<1$, then we have
\be
\lim_{\lambda \rightarrow \infty} \frac{\delta H}{H_{,\lambda}} \rightarrow 0.
\ee
In the single field case this is sufficient to show that all solutions are attractors 
for one another that tend towards convergence in the asymptotic future.
This does not imply the existence 
of a single ``attractor solution," however. This general argument also breaks
down in the presence of large fluctuations \cite{Liddle:1994dx}.
In the multi-field case we must also consider variations in $H$ from isocurvature modes. 
Note that in the collapsing case the same argument shows that $w>1$ solutions
exhibit similar behavior for single field evolution, a well known similarity between
these two scenarios \cite{Erickson:2003zm}.

Another use of the Hamilton-Jacobi form is to generalize the argument for the conservation
of adiabatic curvature modes outside of the horizon beyond the linear level
seen in Eq.~(\ref{linearcons}). Following the arguments presented in
 \cite{Afshordi:2000nr,Geshnizjani:2004tf}, we can define a conserved
 differential perturbation for a single field through the identity
 \be
 \frac{d\zeta}{N dt} = 0 = \frac{d \lambda}{N dt} - H(\phi),
\ee
where $H(\phi)$ is the Hamilton-Jacobi function. Then we can integrate over time to
find
\begin{eqnarray}
\zeta & =  & -\lambda + \int H(\phi) dt \nonumber \\
&= &   -\lambda + \int d \phi \( \frac{d \phi}{d \lambda}\)^{-1} = -\lambda + \int d\phi \({\frac{\partial H_{\rm eq}}{\partial \pi}\(\phi,-2M_P^2 \frac{\partial H}{\partial \phi}\)}\)^{-1}.
\end{eqnarray}
which is nothing other than the integrated form of Eqn.~(\ref{linearcons}), which
correctly behaves as $\zeta \rightarrow \zeta + c$ under time translation. This definition is most useful when computing non-Gaussianity, where we are most interesting in correlators of $\zeta$, since this is the conserved quantity at long wavelengths and is nonlinearly related to the fluctuations of $\phi$.

\subsection{Introducing probability through coarse-graining}
\label{noise}

Let us return to the general equations of motion for a complete field, including
both sub- and super-Hubble modes. Since our aim is to derive the field's
stochastic dynamics, the task at hand is first to characterize the sub-Hubble
modes. We can then coarse-grain over a horizon patch to remove its particular dynamics
and replace them with an averaged ``noise" force acting on the super-Hubble field.
 This noise will characterize
the influence of the sub-Hubble modes on the super-Hubble evolution of the field
as those modes exit the horizon. But first, we must characterize the deterministic
behavior within a horizon, where the ultralocal approximation is invalid. It is thus necessary that we retain
spatial gradients for sub-Hubble modes. The complete system can be written as
$$ \frac{\partial \phi_{\alpha}}{\partial \lambda}=\frac{\delta H_{\rm eq}}{\delta \pi_{\alpha}}, \quad
\frac{\partial \pi_{\alpha}}{\partial \lambda} =-\frac{\delta H_{\rm eq}}{\delta \phi_{\alpha}} - 3 \pi_{\alpha}, \quad {\cal H}_i=0. $$
Here the notation is such that $$\frac{\delta}{\delta \phi}=\frac{\partial}{\partial \phi}-\partial_i \(\frac{\partial}{\partial (\partial_i \phi)} \). $$
Note that we have continued to retain the spatial diffeomorphism constraint $( {\cal H}_i=0)$ since it is important at sub-Hubble scales, but the temporal constraint (${\cal H}=0$) does not enter since we have already explicitly solved for it. We can now divide the fields and their momenta into long and short wavelength
parts.
For any particular field $\phi_{\alpha}$ we may write $$\phi_{\alpha}  = \phi_{\alpha,L} + \phi_{\alpha,S} \, , \quad \pi_{\alpha}  =  \pi_{\alpha,L} + \pi_{\alpha,S}\,.$$
Using this split in the equations of motion, retaining nonlinearity for long wavelengths
but taking the linear approximation for the short wavelengths, we find (suppressing, for notational simplicity, the indices related to field summations)
\begin{eqnarray}
\frac{\partial \phi_L}{\partial \lambda} - \frac{\delta H_{\rm eq}}{\delta \pl} + \( \frac{\partial \phi_S}{\partial \lambda} - \frac{\delta^2 H_{\rm eq}}{\delta \pl^2} \ps -\frac{\delta^2 H_{\rm eq}}{\delta \pl \delta \phl} \phs\) & = & 0, \nonumber \\
\frac{\partial \pi_L}{\partial \lambda} + \frac{\delta H_{\rm eq}}{\delta \phl} + 3 \pl + \( \frac{\partial \pi_S}{\partial \lambda} +  \frac{\delta^2 H_{\rm eq}}{\delta \phl^2} \phs  +\frac{\delta^2 H_{\rm eq}}{\delta \phl \delta \pl} \ps + 3 \ps \) & = & 0, \nonumber \\
{\mathcal H}_i(\phi_L,\pi_L)+\( \frac{\delta {\cal H}_i}{\delta \phi_L} \phi_S+\frac{\delta {\cal H}_i}{\delta \pi_L} \pi_S\)& = & 0.
 \label{longshort}
\end{eqnarray}
With this done, we can subsequently neglect the spatial variation of the long wavelength portion of the field (except in the constraint ${\mathcal H}_i(\phi_L,\pi_L)=0$), and trade $\delta  \rightarrow \partial $, but retain the spatial variation in the sub-Hubble modes. A more careful treatment which continues to use the spatial derivative information at long wavelengths -- which can be useful in calculating non-Gaussianities -- is considered in \cite{Gerry}.

Having now split the field at this level, we need a prescription for precisely how the split is accomplished.
We choose to split each field and its momentum at some particular scale, which will be a constant multiple of the sound horizon for an ordinary field, but which may be more complex in general.
To remain general, we write
\be
\phi_{S} = \int \frac{d^3k}{(2\pi)^3} \Theta(f(\phi_L,\pi_L,\lambda)- k) \delta \phi_k, \quad \pi_{S} = \int \frac{d^3k}{(2\pi)^3} \Theta(f(\phi_L,\pi_L,\lambda)- k) \delta \pi_k
\ee
where we are free choose the function, $f(\phi_L,\pi_L,\lambda)$. We have chosen to use a Heaviside window function (see Ref.~\cite{Winitzki:1999ve} for alternative approaches) because it quite directly retains the Markovian\footnote{Noise is said to be Markovian if its correlator $\langle \eta(\lambda) \eta(\lambda')\rangle$ is proportional to a delta function $\delta(\lambda-\lambda')$.} character of the sub-Hubble noise as considered from super-Hubble scales.

It is clear from the Eqs. (\ref{longshort}), that each $\delta \phi_k$ mode of the field obeys the (linearized) equations of motion; and the
short wavelength part of Eqs. (\ref{longshort}) -- inside the parenthesis -- are exactly these equations of motion.
Hence, the terms in the parenthesis vanish everywhere except when $f = k$, where $k$ is
the horizon scale in the usual case.
 Thus, the derivatives turn $\Theta$ into a $\delta-$function and we can rewrite
 \begin{eqnarray}
\frac{d \phl}{\dl} - \frac{\delta H_{\rm eq}}{\delta \pl} &=& \etap  , \nonumber \\
\frac{d \pl}{\dl} + \frac{\delta H_{\rm eq}}{\delta \phl} + 3 \pl &=& \eta_{\pi} ,
\end{eqnarray}
 where the noise terms, $\etap$ and $\eta_{\pi}$ are given by:
\ba
\label{etaphi}
\etap &=&  -\int \frac{d^3k}{(2\pi)^3} \frac{d  f(\phi_L,\pi_L,\lambda)}{\dl} \delta(f(\phi_L,\pi_L,\lambda) - k) \delta \phi_k\,, \\
\eta_{\pi} &=&  -\int \frac{d^3k}{(2\pi)^3} \frac{d  f(\phi_L,\pi_L,\lambda)}{\dl} \delta(f(\phi_L,\pi_L,\lambda) - k) \delta \pi_k\,.
\ea
Given this definition, one can calculate is the diffusion matrix, $2 D^{(2)}_{A \, B} \delta(\lambda-\lambda') = \langle \eta_A(\lambda) \eta_B(\lambda') \rangle$. 
In general, these two noise terms are independent, since the field has two independent
modes. However, we will typically assume growing mode domination outside of the horizon.
This is equivalent to suppressing one of these modes and implicitly correlates the two noises; 
when this occurs, they cease to be independent. This subtlety does not impact our calculations.
Inserting Eq.~(\ref{etaphi}) and noticing immediately that the correlation between two mode functions with different
momenta will vanish, we can suppress one momentum integration and write
\baray
\langle \etap(\lambda) \etap(\lambda') \rangle &=& \int  \frac{d^3k}{(2\pi)^3}
 \frac{d f(\lambda)}{\dl}  \frac{d f(\lambda')}{\dl'} \delta(f(\phi_L,\pi_L,\lambda) - k)  \delta(f(\phi'_L,\pi'_L,\lambda') - k) \langle \delta \phi_k(\lambda) \delta \phi_k(\lambda') \rangle \\
 &= &  \int  \frac{d^3k}{(2\pi)^3}  \frac{d f(\phi_L,\pi_L,\lambda)}{\dl} \delta(f(\phi_L,\pi_L,\lambda) - k) \delta(\lambda - \lambda') \langle \delta \phi_k(\lambda) \delta \phi_k(\lambda') \rangle \\
 & = &  \frac{4\pi}{(2\pi)^3} \left ( f^2  \frac{d f(\phi_L,\pi_L,\lambda)}{\dl} \) \delta(\lambda - \lambda') \langle \delta \phi_{k=f}(\lambda) \delta \phi_{k=f}(\lambda') \rangle\, ,
\earay
where we have made use of the properties of the delta function several times. Similar expressions can be found for $D^{(2)}_{\phi \, \pi_{\phi}}$ and $D^{(2)}_{\pi_{\phi} \pi_{\phi}}$.

For the super-Hubble modes to be treated classically, it is crucial that the correlators of the noises commute. In practice, this requires
\be
\frac{\langle [\delta \phi_f(\lambda) ,\delta \phi_f(\lambda')] \rangle }{\langle \delta \phi_f(\lambda) \delta \phi_f(\lambda') \rangle} \rightarrow 0,
\ee
i.e. the ratio of the Schwinger commutator function to the Wightman function must tend to zero. This will occur whenever there is a clear definition of a growing mode. In cases where the equation of state $w$ is not varying rapidly, the evolution of a given mode outside the horizon typically splits into a growing part and a decaying part, $\delta \phi_k \sim A e^{g\lambda}+Be^{d \lambda}+\dots$, with $g > d$. In this case, the ratio of the commutator to the Wightman function scales as
\be
\frac{\langle [\delta \phi_f(\lambda) ,\delta \phi_f(\lambda')] \rangle }{\langle \delta \phi_f(\lambda) \delta \phi_f(\lambda') \rangle} \sim e^{(d-g)\Delta\lambda}e^{(d-g)\Delta\lambda'},
\ee
where $\Delta \lambda$ is the number e-foldings through which a given mode has evolved since horizon crossing. Because of this exponential scaling, only a few e-folds are required for the noise to look classical unless $(g-d)$ is particularly small. Of course, any coarse-graining prescription will cease to be valid whenever the modes begin to oscillate. This happens during reheating, for instance, when the mass of the field, not the Hubble constant, dominates the dynamics. However, let us stress that neither slow-roll nor a quasi-de Sitter geometry ($w \approx -1$) is crucial to the stochastic approach {
\it per se}. All that is required is a clearly defined sound horizon (which usually implies $dw/d\lambda \ll 1$), and that the super-Hubble dynamics can be treated ultralocally so that Hubble damping dominates over mass terms. It is true, however, that the quasi-de Sitter case is where these conditions are most easily satisfied.
Let us also stress that there is no problem with treating the modes at super-Hubble scales while they remain slightly quantum, as is the case when it takes a large number of e-foldings for the growing mode to dominate over the decaying modes. In this case, we are just required to use a so-called quantum Langevin equation \cite{Kleinert}.

To proceed we must focus on a specific model, and here we shall consider the general class of single field models whose Lagrangian takes the form $p(X,\phi)$ where the kinetic term given by $X \equiv -\half g^{\mu \nu} \phi,_\mu \phi,_\nu$. We can then appeal to the
general perturbation calculations made in \S8.3 of Ref.~\cite{Mukhanov}. A crucial question is how the perturbations in our approach relate to the gauge invariant variables. As pointed out in Sec.~\ref{scaling}, this relation is in fact straightforward, since we have chosen a gauge in which the volume of spacelike surfaces is used to measure time, then in terms of standard cosmological perturbations this is the gauge in which the spatial metric perturbation $\psi=0$. Thus $\delta \phi_k=\zeta_k \, (\dot{\phi}/H)$. The equation for $\zeta_k$ is of the form
\be
\zeta_k''+2\frac{z'}{z}\zeta'_k+c_s^2 k^2 \zeta_k=0,
\ee
where $'$ denotes conformal time derivative, and $z=a\sqrt{(\epsilon+p)}/(c_sH)$,  and we have
defined a sound speed for the field $$ c_s^2 \equiv \frac{p,_X}{\epsilon,_X}\, , $$ and energy density, $\epsilon$, given by
$ \epsilon \equiv 2 X p,_X - p$. The normalization is set by the Wronskian
\be
z^2 \(\zeta_k' \zeta_k^*-\zeta^{*'}_k\zeta_k\)=-2i .
\ee
At super-Hubble scales the solution for the positive frequency mode looks like
\be
\zeta_k^+ =A_k +B_k \int \frac{1}{z^2} \, d \eta,
\ee
which represent the growing ($A_k$) and decaying ($B_k$) modes. To determine these approximately we may use the horizon crossing approximation, which is valid whenever the sound horizon is well defined ($d \ln (c_s^{-1} aH)/d\lambda \ll 1$).
The normalization of $\delta \phi_k$ is given by $$\delta \phi_k^+ = \frac{\eth}{a\sqrt{\epsilon,_X}} \frac{1}{\sqrt{2 f c_s}}\, , $$ where $\eth$ is a fudge factor of order $1$ that reflects the inaccuracy of the horizon crossing approximation. We can fix this fudge factor by comparing with exactly solvable models \cite{Kinney}, and in particular those that give scale-invariant spectra. Taking $f =\xi aH/c_s$, with $\xi \ll 1$ to ensure growing mode dominance, we obtain
\be
\(D^{(2)}_{\phi \, \phi}\, , D^{(2)}_{\phi \, \pi_{\phi}}, D^{(2)}_{\pi_{\phi}\pi_{\phi}}\)= \(\frac{M_P^2H^2}{8\pi^2c_sp_{,X}} ,\frac{\dot{\pi}}{\dot{\phi}}\frac{M_P^2H^2}{8\pi^2c_sp_{,X}}, \(\frac{\dot{\pi}}{\dot{\phi}}\)^2\frac{M_P^2H^2}{8\pi^2c_sp_{,X}}\)
\ee
Given the Langevin equation, we can infer an equation for the evolution of the probability density on phase space using Appendix B, which takes the form of a Fokker-Planck equation. Here, we write the Fokker-Planck equation as a current conservation equation
\be
\label{conservationequation}
\frac{\partial \rho}{\partial \lambda}= -\frac{\partial  {\cal J}_A}{\partial \phi_A},
\ee
where ${\cal J}_A$ is the current and we are using a notation so that $\phi_A=(\phi_{\alpha},\pi_{\alpha})$ represents any canonical coordinate. For a diffusion matrix given by $D_{AB}$, the currents take the general form
\be
{\cal J}_A=\(\Omega_{AB}\frac{\partial H_{\rm eq}}{\partial \phi_A}-\gamma_{A}{}^{B}\phi_B\)\rho -D_{AB}\frac{\partial}{\partial \phi_A}\rho,
\ee
where $\Omega_{AB}$ is the symplectic form on phase space (which can easily be determined
from any particular model's equations of motion) and $\gamma_{AB}$ is the damping term (for all inflationary systems, $\gamma_{\pi_a}{}^{\pi_b}=3\delta_{ab}$ is the only nonzero
damping term).
In particular, for a single scalar field, we have
\be
\( {\cal J}_{\phi},{\cal J}_{\pi_{\phi}}\)= \(\frac{\partial H_{\rm eq}}{\partial \pi_{\phi}} -D_{\phi \, \phi} \frac{\partial \rho}{\partial \phi}-D_{\phi \, \pi_{\phi}} \frac{\partial \rho}{\partial \pi_{\phi}} ,
 -\frac{\partial H_{\rm eq}}{\partial \phi} -3\pi_{\phi}-D_{\pi_{\phi}\phi} \frac{\partial \rho}{\partial \phi}-D_{\pi_{\phi}\pi_{\phi}} \frac{\partial \rho}{\partial \pi_{\phi}} \).
\ee
In writing these relations we have made a specific choice of operator ordering, which is connected with the precise measure in the path integral of Appendix B. It is common to consider the more general ordering of the diffusion term
\be
D_{AB}\frac{\partial}{\partial \phi_A}\rho\rightarrow D_{AC}\, (Q^{-1})^{CD}\frac{\partial}{\partial \phi_A}
\(Q_{DB}\rho\),
\ee
where for instance $(Q)_{AB}=D^{\beta}_{AB}$ for $0\le \beta \le 1$. This typically gives a sub-leading effect to the dynamics, so we choose $\beta=0$. The resulting ordering of terms is known
as the Ito ordering; we further note that ref. \cite{Vilenkin:1999kd} argues
that this prescription is the natural one for stochastic inflation.

\subsubsection{One-noise coarse-graining}

In the one-noise case the procedure is very similar. As before we split $\phi_{\alpha}=\phi_{\alpha,L}+\phi_{\alpha,S}$, but now rather than splitting $\pi$, we define $\pi_{\alpha,L}$ in terms of $\phi_{\alpha,L}$ using the long wavelength equations. In other words the equation
\be
\frac{\partial\phi_L}{\partial \lambda} - \frac{\delta H_{\rm eq}}{\delta \pl}=0,
\ee
remains exact without noise, so that $\eta_{\phi}=0$. We define
\be
\phi_{S} = \int \frac{d^3k}{(2\pi)^3} W(f(\phi_L,\pi_L,\lambda)- k) \delta \phi_k,
\ee
where the window function now must be chosen differently to maintain a Markovian noise. As before the second equation includes a noise term
\be
\frac{\partial \pi_L}{\partial \lambda} + \frac{\delta H_{\rm eq}}{\delta \phl} + 3 \pl = \eta_{\pi},
\ee
so that the only nonzero Diffusion term is $D^{(1)}_{\pi\pi}=\frac{1}{2}\langle \eta_{\pi}\eta_{\pi}\rangle$.

\subsection{Systems satisfying Detailed Balance}
\label{detailedbalance}

The existence of an effective time-translation symmetry allows us to consider models which satisfy the principle of detailed balance. 
We choose to explore detailed balance because using it will give us a more intuitive 
understanding of de Sitter entropy as an equilibrium in the evolution of gravitationally
coupled scalar field dynamics. The concept of de Sitter entropy is important enough
that its appearance here, emerging from an approach quite different from its usual derivation,
is worth comment.  
In the present context, the principle of detailed balance
states that an equilibrium state should exist in which the loss of `energy' (where energy is $H_{\rm eq}$) via damping is on
average exactly compensated by the gain in energy from the small scale stochastic noise fluctuations. 
More precisely: for a discrete system with probabilities $p_n$, detailed balance states
that given the master equation \be
\frac{dp_n}{d\lambda}=\sum_m \(-r_{n\rightarrow m}p_n +  r_{m\rightarrow n}p_m\),\ee where $r_{n \rightarrow m}$ denotes the transition rates between states $n$ and $m$, in equilibrium we should have
$r_{n\rightarrow m}p_n =r_{m\rightarrow n}p_m$. Generalizing to a continuous system with probability distribution $\rho$,
we replace the above with the probability current conservation equation (\ref{conservationequation}).
For a system that exhibits overall time translation invariance, the current may always be split into a reversible (even) and irreversible (odd) part \be {\cal J}_A={\cal J}_A^{\rm rev}+ {\cal J}_A^{\rm irrev},\ee where time reversal symmetry $T$ ($\lambda \rightarrow -\lambda$) acts as $$
T (\phi_A,{\cal J}_A^{\rm rev},{\cal J}_A^{\rm irrev}) =  (\epsilon_A\phi_A,-\epsilon_A{\cal J}_A^{\rm rev},\epsilon_A{\cal J}_A^{\rm irrev}) ,$$
and, again, $\epsilon_{A}=\pm 1$ is the intrinsic time parity of the field $\phi_A$. Note that $\pi_{\alpha}$ has the opposite
time parity to $\phi_{\alpha}$ by definition.
For a continuous system of even and odd variables, detailed balance amounts to the statement that, in equilibrium\footnote{For a detailed discussion of the reason behind this the reader is encouraged to consult section 6.4 in Ref.~\cite{Risken:1989fp}.}, \be {\cal J}^{\rm irrev}_A=0 \quad {\rm and} \quad
\frac{\partial {\cal J}^{\rm rev}_A}{\partial \phi_A}=0. \ee
Schematically, we can see how this works out: the damping term and the diffusive noise contributions are both
irreversible contributions.  Consequently, for a detailed-balance equilibrium to form, they must exactly cancel each other out.

To see how detailed balance can be implemented more concretely, let us focus on a single field system described by the conjugate
variables $\phi,\pi_{\phi}$. The reversible current is that determined by the reversible, or ``equilibrium," Hamiltonian, $H_{\rm eq}$: \be
({\cal J}^{\rm rev}_{\phi},{\cal J}^{\rm rev}_{\pi_{\phi}})=(\frac{d {\phi}_{\rm eq}}{d\lambda}\rho,\frac{d {\pi_{\phi \rm eq}}}{d\lambda}\rho)=\(\frac{\partial H_{\rm eq}}{\partial
\pi_{\phi}}\rho,-\frac{\partial H_{\rm eq}}{\partial \phi}\rho\) .\ee The irreversible current contains two contributions, one from the
damping current \be ({\cal J}^{\rm damp}_\phi,{\cal J}^{\rm damp}_{\pi_{\phi}})= (\frac{d {\phi}_{\rm damp}}{d\lambda}\rho,\frac{d {\pi_{\phi,{\rm damp}}}}{d\lambda}\rho)=(0,-3\pi_{\phi} \rho),\ee and a
diffusive noise current that we expect to be of the form\footnote{This is the correct form of the diffusion current if the stochastic noise is Markovian and Gaussian according to the arguments given in Appendix B.} \be {\cal J}^{\rm diff}_A=-D_{AB}\frac{\partial \rho}{\partial \phi_B}. \ee
In the one-noise approach, the diffusive current takes the simpler form \be
({\cal J}^{\rm diff}_{\phi},{\cal J}^{\rm diff}_{\pi_{\phi}})=(0,-D^{(1)}_{\pi_{\phi}\pi_{\phi}}\frac{\partial \rho}{\partial \pi_{\phi}}). \ee

We are now ready to determine the consequences of assuming detailed balance. The general definition of an equilibrium state is $\partial^A {\cal J}^{\rm, rev}_A=\{\rho,H_{\rm eq}\}=0$. This implies that $\rho_{\rm eq}=F(H_{\rm eq})$. The detailed balance requirement, ${\cal J}^{\rm irrev}=0$, implies $-3\pi_{\phi}\rho-D^{(1)}_{\pi_{\phi}\pi_{\phi}} \frac{\partial \rho_{\rm eq}}{\partial
\pi_{\phi}}=0$, which fixes the diffusion parameter to be \be D^{(1)}_{\pi_{\phi}\pi_{\phi}}=-\frac{3\pi_{\phi}}{\frac{\partial H_{\rm eq}}{\partial \pi_{\phi}}
\frac{\partial \ln F(H_{\rm eq})}{\partial H_{\rm eq}}} .\ee Thus detailed balance uniquely fixes the form of the Fokker-Planck equation down to
the specification of a single function, the equilibrium distribution $\rho_{\rm eq}=F(H_{\rm eq})$. Once we have reached this state,
though, we can take guidance from the fluctuation theorem \cite{FT}. It tells us that, in equilibrium, the probability must be given by $\rho_{\rm eq}=\exp{S_{\rm ent}}$,
where $S_{\rm ent}$ is the entropy of the equilibrium configuration. Now, in any system in which gravity is given by the Einstein action we know that
that $H_{\rm eq}=6M_{P}^2 H$ where $H$ is the Hubble parameter. Thus, all that is required for a complete specification of the system
is a definition of the entropy of the system as a function of the Hubble parameter. There is, of course, a natural definition of this entropy: the
horizon entropy in a de Sitter space background, $S_{\rm ent}=8\pi^2M_P^2H^{-2}$. Thus, the natural choice of equilibrium distribution is
$\rho_{\rm eq}=\exp(8\pi^2M_P^2(6M_{P}^2)^2H_{\rm eq}^{-2}).$  This fixes the diffusion parameter to be \be D^{(1)}_{\pi_{\phi}\pi_{\phi}}=\frac{3\pi_{\phi} H_{\rm eq}^3}{16\pi^2M_P^2(6M_{P}^2)^2\frac{\partial
H_{\rm eq}}{\partial \pi_{\phi}}}. \ee
Comparing with the result of the the coarse-graining calculation, this agrees only if $c_s=1$. Thus detailed balance is exactly satisfied for minimally coupled models, but in general is only approximately true in the region of phase space for which $c_s\approx 1$. Nonetheless, it is remarkable that
the de Sitter entropy prescription -- $\rho \sim e^S$ -- drops out so naturally from this line of argument.
Viewed in this light, the entropy of de Sitter space is a fixed point of a Fokker-Planck evolution
that can begin very far from equilibrium. Since equilibria and maximal entropy are naturally
related thermodynamic concepts, this approach to de Sitter entropy is considerably 
easier to grasp, intuitively, than the usual coarse-graining derivation.

\subsubsection{Generalizing to Two-Noises}
\label{twonoise1}

In the previous section, we used detailed balance and the assumption of a single noise acting on
 the field's
momentum to derive an ansatz equilibrium distribution function. It turns out to often be nicer
to work in a more general system where we retain noise-driven diffusion in both the $\phi$ and $\pi_{\phi}$ directions, so we will generalize our results to this case.

As we showed in section \ref{scaling}, the existence of a scaling symmetry guarantees that a solution to the perturbed equations of motion exists that satisfies the adiabaticity condition
\be
\frac{\delta \phii}{\phil} = \frac{\delta \pii}{\pil}.
\ee
Assuming that this solution is the dominant growing mode, the diffusion terms appearing in the Fokker-Planck
equation will be related by
\be
\(D^{(2)}_{\phi_{\alpha}\,\phi_{\beta}},D^{(2)}_{\phi_{\alpha}\,\pi_{\beta}},D^{(2)}_{\pi_{\alpha} \, \pi_{\beta}} \)=\(\frac{d\phi_{\alpha}}{d\lambda}\frac{d\phi_{\beta}}{d\lambda}, \frac{d\phi_{\alpha}}{d\lambda}\frac{d\pi_{\beta}}{d\lambda},\frac{d\pi_{\alpha}}{d\lambda}\frac{d\pi_{\beta}}{d\lambda}\)Z,
\ee
Now, we shall check to see that the currents in the $\phi$ and $\pi_{\phi}$ directions vanish in equilibrium. We find
\be
{\cal J}_{\phii}  =  \dHdpi \rho -  \( D^{(2)}_{\phii \phij} \frac{\partial \rho}{\partial \phi_{\beta}} + D^{(2)}_{\phii \pij} \frac{\partial \rho}{\partial \pi_{\beta}}  \)
 =  \dHdpi \rho-D^{(2)}_{\phii \phij}   \(\frac{\partial \rho}{\partial \phi_{\beta}} +  \left (\frac{\pil}{\phil} \right )  \frac{\partial \rho}{\partial \pi_{\beta}}  \). \label{diffop}
\ee
Following the same reasoning for ${\cal J}_{\pii}$, we find (under the assumption of adiabaticity)
\be
{\cal J}_{\pii}= \frac{\pil}{\phil}  {\cal J}_{\phii}.
\ee
Thus, there must exist an equilibrium state in which ${\cal J}_A=0$ identically! This is the statement of detailed balance in a system whose degrees of freedom are neither even nor odd under
time parity.

The vanishing of the current implies that $F$ satisfies
\be
Z^{-1}=\frac{d}{d \lambda}\ln(\rho_{\rm eq})=\(\frac{d\phi_{\alpha}}{d\lambda}\frac{\partial}{\partial \phi_{\alpha}}+\frac{d\pi_{\alpha}}{d\lambda}\frac{\partial}{\partial \pi_{\alpha}}\) \ln(\rho_{\rm eq}).
\ee
Since this is only a single differential equation, modulo obstructions, one can basically always find equilibrium solutions provided they are normalizable.
This is not of the same form as in the one-noise case, but it is easy to demonstrate that
\be
D_{AB}\frac{\partial H_{\rm eq}}{\partial \phi_A}\frac{\partial H_{\rm eq}}{\partial \phi_B}=D^{(1)}_{AB}\frac{\partial H_{\rm eq}}{\partial \phi_A}\frac{\partial H_{\rm eq}}{\partial \phi_B}=D^{(2)}_{AB}\frac{\partial H_{\rm eq}}{\partial \phi_A}\frac{\partial H_{\rm eq}}{\partial \phi_B},
\ee
is identical in both the one- and two-noise cases. In the two-noise case, any equilibrium distribution related to the diffusion constant in this way is consistent with detailed balance. However, in the two-noise case there is no a priori reason that $\rho_{\rm eq}$ should only be a function of $H_{\rm eq}$, and in particular we shall find that for $c_s \neq 1$ it cannot retain this simple form in general.

\subsection{DBI Inflation}

\label{DBIinflation}

A natural application of our present results is to give a stochastic description of DBI inflation. In this case, inflation is driven by a scalar field with a non-canonical kinetic term.
%Inflation can occur in these theories outside of the usual slow-roll limit.
Our previous discussions were sufficiently general to encompass this case.
To specialize to DBI requires only a quick calculation of
 $H^{DBI}_{\rm eq}(\phi,\pi_{\phi})$ to describe the full dynamics. We will make an
explicit calculation to demonstrate this.

The action for DBI inflation is
\be
S=\int d^4 x \sqrt{-g} \( \frac{M_{P}^2}{2}R-\frac{1}{f(\phi)}\sqrt{1+f(\phi) (\partial \phi)^2} + \frac{1}{f(\phi)}-V(\phi)\)
\ee
We have absorbed any conformal couplings into a field redefinition and so are working in Einstein frame. The mini-superspace form of this action is
\be
S=\int dt N e^{3\lambda}\( -3M_{P}^2 \frac{\dot{\lambda}^2}{N^2} -f^{-1}\sqrt{1-f\frac{\dot{\phi}^2}{N^2}} + f^{-1}-V\),
\ee
and so after solving for $N$ and substituting back in as before we find the equilibrium Hamiltonian,
$H_{\rm eq} = 6 M_{P}^2 H $, given by
\be
H^2_{\rm eq} = 12M_{P}^2 \( f^{-1}\sqrt{1+f \pi_{\phi}^2} - f^{-1}+V \).
\ee
So the reversible currents for DBI inflation are
\begin{eqnarray}
{\cal J}^{\rm rev}_{\phi} &  = & \frac{6M_P^2}{H_{\rm eq}} \frac{\pi_\phi}{\sqrt{1+ f \pi_\phi^2}} \rho, \\
{\cal J}^{\rm rev}_{\pi_{\phi}} &  = & - \frac{6M_P^2}{ H_{\rm eq}}\left \{ \frac{\partial \ln f}{\partial \phi} \( \frac{\half \pi_\phi^2}{\sqrt{1 + f \pi_\phi^2}} - f^{-1} \( \sqrt{1+f\pi_\phi^2} - 1 \) \)
 - \frac{\partial V}{\partial \phi} \right \} \rho.
 \end{eqnarray}
When we solve these equations numerically in Sec.~\ref{numerics}, we will
work in the one-noise formulation. In that case, there is only one diffusion parameter, $D_{\pi_{\phi}\pi_{\phi}}$.
An explicit calculation gives the DBI diffusion parameter to be
\be
D^{(1)}_{\pi_{\phi}\pi_{\phi}}=\frac{9 M_P^2}{8\pi^2 c_s^2}H^4.
\ee

\subsubsection{Comparison with previous work}

In the previous work described in Ref.~\cite{Chen:2006hs} a stochastic equation for DBI was derived using the Hamilton-Jacobi approach. Here we show that our two-noise formulation with the assumption of adiabaticity is completely equivalent to the Hamilton-Jacobi formulation.
To prove this, let us begin with the two-noise Langevin system for a single field
\be
\frac{d\phi_\alpha}{d\lambda}=\frac{\partial H_{\rm eq}}{\partial \pi_{\alpha}}+\eta_{\phi_\alpha} ,\quad \frac{d\pi_{\alpha}}{d\lambda}=-\frac{\partial H_{\rm eq}}{\partial \phi_{\alpha}}+\eta_{\pi_{\alpha}}.
\ee
As in Sec.~\ref{Hamilton-Jacobi} we perform a change of variables $\pi_{{\alpha}} =f_{\alpha}(\phi_{\beta},J_{\beta})$ so that we have
$$
\frac{d\pi_{{\alpha}}}{d\lambda}-f_{,\phi_{\alpha}}\frac{d\phi_{\alpha}}{d\lambda}=f_{\alpha,J_{\beta}}\frac{dJ_{\beta}}{d\lambda}.
$$
If the noise is adiabatic, then $$\frac{\frac{d\phi_{\alpha}}{d\lambda}}{\frac{d\pi_{{\alpha}}}{d\lambda}}=\frac{\eta_{\phi_\alpha}}{\eta_{\pi_{\alpha}}}=\frac{\( \frac{\partial H_{\rm eq}}{\partial \pi_{{\alpha}}}\)}{\(-3\pi_{\phi_{\alpha}}-\frac{\partial H_{\rm eq}}{\partial \phi_{\alpha}}\)}$$ for both the classical and quantum parts, so the same choice of function $f_{\alpha}=-2M_P^2\frac{\partial H}{\partial \phi_{\alpha}}$ as before implies that the equation
$$\frac{dJ_{\alpha}}{d\lambda}=0, $$ is true both classically and with noise included. Here we see that the assumption of adiabaticity is a crucial ingredient, since $J_{\alpha}$ represents the decaying (non-adiabatic) mode of the perturbations and the presence of any non-adiabaticity would give kicks to $J_{\alpha}$ even if classically it is conserved. Explicitly then, for single-field DBI the remaining stochastic equation is
$$
\frac{d\phi}{d\lambda} =-\frac{2M_P^2 c_s H_{,\phi}}{H}+\eta_{\phi},
$$
with $\langle \eta_{\phi}\eta_{\phi}\rangle =2D^{(2)}_{\phi \, \phi}=H^2/(4\pi^2)$, and $$ c_s=\frac{1}{\sqrt{1+4M_P^4 f{H_{,\phi}}^2}}.$$
This is equivalent to the equation in \cite{Chen:2006hs} converted to e-folding time.
The choice of which formalism to use is one of convenience. It is generally difficult to solve the Hamilton-Jacobi equation and retain the dependence on $J$; for this reason,  we shall concentrate on the phase space approach for our
numerical solutions. A second reason for concentrating on the one-noise approach for numerics, is that as described in the introduction, the DBI speed limit $d\phi/dt \le 1/\sqrt{f}$ is in general violated in the full two-noise/Hamilton-Jacobi stochastic dynamics.
However, the Hamilton-Jacobi version has the advantage of allowing us to give explicit equilibrium solutions (at least formal ones, written in terms of $H(\phi,J)$), and is also useful in understanding tunneling.

\subsubsection{Equilibrium solutions}

The equivalence between the two-noise and Hamilton-Jacobi approaches allows us to find simple equilibrium solutions which satisfy detailed balance. The Fokker-Planck equation (in which $J$ derivatives drop out) in Hamilton-Jacobi form is
\be
\frac{\partial \rho}{\partial \lambda}+\frac{\partial}{\partial \phi} \(-\frac{2M_P^2 c_s H_{,\phi}}{H}\rho- \frac{H^2}{8\pi^2} \frac{\partial}{\partial \phi} \rho\)=0.
\ee
This has explicit solutions
\be
\rho_{\rm eq}=\exp \( -\int^{\phi} d\phi \frac{16\pi^2M_P^2 c_s H_{,\phi}}{H^3} \).
\ee
For $c_s=1$ this gives the standard $e^{S_{\rm ent}}$ solutions.
Strictly speaking we should check that these solutions are normalizable, but this is essentially guaranteed if $H$ has a nonzero minimum, as is necessary to continue to coarse-grain over Hubble size volumes, and if we impose an upper boundary in $\phi$ space, as is required to ensure the validity of effective field theory.
We stress again that this is in fact a 1-parameter family of solutions since $H$ has an implicit dependence on $J$. This freedom is a reflection of the fact that because there is no noise acting in the $J$ direction, there is nothing to bring the $J$ dependence to equilibrium. In practice adiabaticity is not exactly true, and so we can imagine that a more unique solution would arise, but it would take much longer for an equilibrium to set in for the dynamics of $J$ in comparison with those of $\phi$.

In Sec.~\ref{tunneling}, we shall make use of these equilibrium solutions to understand tunneling. For now we merely point out that, whilst the peak of the equilibrium distribution is where $H_{,\phi}=0$ where $c_s=1$, the width of the distribution will be significantly wider if $c_s \ll 1$ away from this peak. In particular in the ultra-relativistic limit we have
\be
\rho_{\rm eq} = \exp \( -\int^{\phi} d\phi \frac{8\pi^2(3M_P^2)^{3/2}}{\sqrt{f}V^{3/2}}\),
\ee
away from the maximum. Note that for $f \sim \phi^{-4}$ and $V\sim \phi^2$ we have $\ln \rho_{\rm eq} \propto \ln \phi$. Thus for smaller D-brane tensions, the distribution will be broader.

\section{Properties of the Fokker-Planck Equation}

\label{properties}

\subsection{Condition for Classicality versus Eternal Inflation}

Stochastic noise arises fundamentally from small scale quantum fluctuations which appear classical at super-Hubble scales. The large scale evolution is predominantly classical as long as the average effect of the noise is smaller than the classical evolution. The regime in which the noise becomes comparable to the damping is the regime of onset of eternal inflation. There, quantum corrections are large enough
to overwhelm the classical trajectory, leading to a potentially unlimited period of inflation.

We can give a simple prescription for the onset of the eternally inflating regime by looking at the Langevin equation for the Hamiltonian (or equivalently Hubble constant). Its evolution is given by
\be
\frac{dH_{\rm eq}}{d\lambda}=-\gamma_A^B \phi_B \frac{\partial H_{\rm eq}}{\partial \phi_A}+\eta_A \frac{\partial H_{\rm eq}}{\partial \phi_A},
\ee
where $\gamma_A^B$ is a generalization of the damping parameter. The rate of energy loss due to damping (the classical part) dominates the noise (the quantum part) whenever
\be
\(\gamma_A^B \phi_B \frac{\partial H_{\rm eq}}{\partial \phi_A}\)^2 \ge D_{AB}  \frac{\partial H_{\rm eq}}{\partial \phi_A} \frac{\partial H_{\rm eq}}{\partial \phi_B}.
\ee
In either the one or two-noise approach -- we have
\be
\gamma_A^B \phi_B \frac{\partial H_{\rm eq}}{\partial \phi_A}=D_{AB}  \frac{\partial H_{\rm eq}}{\partial \phi_A} \frac{\partial H_{\rm eq}}{\partial \phi_B} c_s \frac{dS_{\rm ent}}{dH_{\rm eq}},
\ee
where the de Sitter entropy is $S_{\rm ent}=8\pi^2M_{P}^2H^{-2}=8\pi^2M_P^2(6M_{P}^2)^2/H_{\rm eq}^2$, and so the inequality takes the form
\be
\frac{dS_{\rm ent}}{dH_{\rm eq}} \(\gamma_A^B \phi_B \frac{\partial H_{\rm eq}}{\partial \phi_A}\) \ge \frac{1}{c_s},
\ee
which is essentially equivalent to
\be
\frac{dS_{\rm ent}}{d\lambda} \ge \frac{1}{c_s},
\ee
reducing for minimal kinetic term models to the intriguing result $\frac{dS_{\rm ent}}{d\lambda} \ge {1}$.
For the evolution of a system to remain classical, its change in entropy per e-fold must be greater than unity or even larger. This result is consistent with that derived in \cite{ArkaniHamed:2007ky}, where more general arguments were given based on the null energy condition. It demonstrates that it is much easier to achieve eternal inflation for small values of $c_s$, as seen also in \cite{Helmer:2006tz}. In \cite{ArkaniHamed:2007ky} this was used as an argument to say that there are a maximum number of e-folds possible before reaching an eternally inflating regime, given by $N_{\rm max}=c_s S_{\rm ent}$.

For the specific case of DBI inflation this condition amounts to
\be
fH^4 \le 1,
\ee
which in the regimes of most interest is $ fV^2/M_{P}^4 \le 1$. In the simple scenario considered in Ref.~\cite{Chen:2006hs}, $f \sim 1/\phi^4$ and $V \sim \phi^2$ and so this ratio is constant. If it is fixed by the requirement of giving phenomenologically acceptable inflation, then its magnitude is $10^{-10}$ and so these authors concluded that it was impossible to achieve eternal inflation. It is clear that this result arises from an accidental relationship between $f$ and $V$ which is highly model dependent. It seems unlikely that as new models are developed this coincidence shall remain. In Sec.~\ref{numerics} we get around this by assuming that the potential is modified, as is expected, at large field values, and in particular take $V \sim \phi^4$. With such a choice, if $\phi$ increases by a factor of $300$, this is sufficient to cancel the $10^{-10}$ factor and reach an eternally inflating regime.

\subsection{Large Damping versus ultra-relativistic Limit}

The system is strongly damped when the damping term, $-3\pi_{\phi}$,
dominates over time derivative of the momentum, $d\pi_{\phi}/d\lambda$. This limit is most straightforward to understand in the two-noise/Hamilton-Jacobi approach.
Defining a damping parameter $\gamma_0=3$, the momentum is given by $\pi_{\phi}=-\frac{6M_P^2}{\gamma_0} \frac{\partial H_{\rm eq}}{\partial \phi}$.  Taking the large $\gamma_0$ limit is equivalent to sending $\pi_{\phi}$ to zero for fixed $H_{\rm eq}$.

We can then expand in powers of $1/\gamma_0$, so that $H^2\approx V/(3M_P^2)$ and the Langevin equation becomes
\be
\frac{d\phi}{d\lambda}=-M_P^2\frac{V_{,\phi}}{V}+\eta_{\phi},
\ee
where now $D_{\phi \, \phi}=\frac{V}{24\pi^2M_{P}^2}$. Using the standard rules described in Appendix B, this gives rise to the Fokker-Planck equation,
\be
\frac{\partial \rho}{\partial \lambda}+\frac{\partial}{\partial \phi}\(-M_P^2\frac{V_{,\phi}}{V}\rho-\frac{V}{24\pi^2 M_{P}^2}\frac{\partial \rho}{\partial \phi} \)=0,
\ee
which is nothing but the familiar Starobinsky equation expressed in terms of e-folding time.

On the other hand, the ultra-relativistic limit for DBI is the limit in which the tension of the D-brane becomes small, i.e. $f(\phi)$ becomes very large. In this limit we also have $H^2\approx V/(3M_P^2)$, since the kinetic term, for fixed $\pi_{\phi}$, tends to zero in this case. The ultra-relativistic Langevin equation is of the form
\be
\frac{d\phi}{d\lambda}=-{\rm sign}(V_{,\phi})\sqrt{\frac{3M_P^2}{f V}}+\eta_{\phi},
\ee
with the same noise normalization as before. It is amusing to note that the highly damped and ultra-relativistic limits behave the same if $f=\frac{3V}{M_P^2V_{,\phi}^2}$. In the case of the models considered in Sec.~\ref{numerics}, this would be achieved with a pure $\frac{1}{2}m^2 \phi^2$ potential whenever $\phi \ll b$ (since $f(\phi) = \Lambda/(\phi^2+b^2)^2$), provided that $\Lambda=3b^4/(3m^2M_P^2)$. Whilst this is not necessarily a realistic parameter regime, it nicely illustrates the fact that one can achieve nearly identical behavior in two very different regimes of phase space, or for two very different models.

\subsection{From e-folding to proper time}

Although we have made use of e-folding time throughout, it is often convenient and more familiar to work with proper time. It is in fact straightforward to perform the conversion, since $dt=H^{-1}d\lambda=6M_P^2 d\lambda/H_{\rm eq}$. The equilibrium Hamiltonian governing evolution in proper time is given by $H^t_{\rm eq}=H_{\rm eq}^2/(12M_P^2)$, whereas the diffusion constants are redefined as
\be
D^t_{AB}=\( \frac{H_{\rm eq}}{6M_P^2}\) D_{AB},
\ee
so the proper time Fokker-Planck equation is given by $
\frac{\partial \rho}{\partial t}= -\frac{\partial  {\cal J}^t_A}{\partial \phi_A}$,
where ${\cal J}^t_A$ is the current
\be
{\cal J}^t_A=\(\Omega_{AB}\frac{\partial H^t_{\rm eq}}{\partial \phi_A}-\gamma_{A}{}^{B}\phi_B\frac{H_{\rm eq}}{6M_P^2}\)\rho -D^t_{AB}\frac{\partial}{\partial \phi_A}\rho.
\ee
Both of these probability distributions are conserved with respect to the same phase space measure
\be
\int d\phi_A \, \rho=\int \int d \phi_{\alpha} d\pi_{\alpha} \, \rho =1.
\ee

\subsection{Entropy and Irreversibility}

The Fokker-Planck equation is an irreversible equation. This means
that whenever equilibrium solutions exist and are reachable, all solutions to the Fokker-Planck
equation will approach them. Whether this solution is unique depends on the precise form of the equation as well as the specification of boundary conditions. For instance, when
we impose no flux boundary conditions and there exists a detailed balance equilibrium
solution, this solution will be a consistent final state for the system. In any event, the
equation's irreversibility can
be demonstrated by proving the analogue of Boltzman's H-theorem. Defining the information entropy \be {\mathcal S}=-\int \int d \phi_{\alpha} d \pi_{\alpha}
\rho \ln (\rho/\rho_{\rm eq})=-\int d \phi_A
\rho \ln (\rho/\rho_{\rm eq}), \ee where $\rho_{\rm eq}$ is any equilibrium solution of the Fokker-Planck equation, then it is a textbook result \cite{Risken:1989fp}
that \be \frac{d \mathcal S}{d\lambda} = -\int d \phi_A \(  \frac{\partial \rho}{\partial \lambda}(1+\ln
(\rho/\rho_{\rm eq}))-\frac{\rho}{\rho_{\rm eq}^2}\frac{\partial \rho_{\rm eq}}{\partial \lambda}\) \ee \be = \int d \phi_A \frac{D_{AB}}{\rho}
\left(\frac{\partial \rho}{\partial \phi_A}-\frac{\rho}{\rho_{\rm eq}}\frac{\partial \rho_{\rm eq}}{\partial \phi_A} \right) \left(\frac{\partial \rho}{\partial \phi_B}-\frac{\rho}{\rho_{\rm eq}}\frac{\partial \rho_{\rm eq}}{\partial \phi_B} \right)\, \ge 0,\ee
where the inequality follows from the fact that $D_{AB}$ is a positive semi-definite matrix which follows naturally from its definition as $D_{AB}=\frac{1}{2}\langle \eta_A \eta_B \rangle$. In practice the entropy increases monotonically until $\rho=\rho_{\rm eq}$.
It is straightforward to prove (for instance using the path integral formulation in Appendix B) from the Fokker-Planck equation that
\be
\frac{|P(A\rightarrow B;\lambda)|}{|P(B\rightarrow A;\lambda)|} = e^{\Delta \ln \rho_{\rm eq}},
\ee
where $P(A\rightarrow B;\lambda)$ is the probability (strictly speaking, probability density) to go from a state $A$ to a state $B$ in a time $\lambda$. Both of the above statements show that there is an irreversible flow towards the equilibrium configuration, which is a maximal entropy (in the sense of ${\cal S}$) state. The fluctuation theorem \cite{FT} implies that $\rho_{\rm eq}=\exp{S_{\rm ent}}$ where $S_{\rm ent}$ is the entropy of the system and in the case $c_s=1$ at least this is certainly consistent.

\subsection{Measure Issues}
One of the great hopes for the stochastic inflationary method is that it will give insight into what a natural measure on inflationary
phase space should, without which it is very difficult to assess the predictivity or robustness of any of the numerous inflationary models
that have been proposed. The Fokker-Planck equation that we have constructed is a probability evolution equation with suggestive
equilibrium solutions, but these solutions do not, as they stand, constitute a measure. Indeed, in the usual case with $c_s=1$ this certainly does not provide a suitable measure for the initial conditions for inflation since it gives a probability distribution which is heavily peaked at the bottom of the lowest potential well, whereas to start inflation we need to start high up. Furthermore once an equilibrium has been reached, the probability of jumping up to the top of the hill is exponentially small (zero in the limit of true equilibrium).
It is interesting to note that this old problem can be somewhat ameliorated by relativistic effects, since these typically broaden the distribution.

\subsubsection{Volume Weighting}

In formulating models of eternal inflation, it is common to weight solutions by their physical volume. The physical reason is that the stochastic probability distribution we have considered so far describes physics averaged over one Hubble volume. However, as space expands the number of Hubble volumes increases and so a `superobserver' would multiply the probability measure by the number of such Hubble volumes $\propto e^{3 \lambda}H^{3}$.

The implications of this are highly gauge dependent (see Ref.~\cite{Winitzki:2005ya} for a discussion on this). In fact, if we use e-folds as are natural time variable, i.e. surfaces of constant volume defining time, then there is no meaningful effect of volume weighting. Consequently, it has been notoriously difficult to define volume weighting in a gauge invariant manner. This lack of gauge invariance is tied to the fact that no observer can actually observe all these different Hubble patches, and so there is no natural frame in which to address this question. We have nothing new to add to this issue here, and refer the reader two recent approaches \cite{Hartle:2007gi} and \cite{Linde:2007nm} at tackling this thorny question.

The standard approach is to use proper time slices, and weight by $\exp{(3\int^t H dt)}$. This can be incorporated into the proper time Fokker-Planck equation replacing it with
\be
\frac{\partial \rho}{\partial t}=3(H-\langle H \rangle)\rho-\partial_A {\cal J}^A_t,
\label{volumeweightedFP}
\ee
where $\langle H \rangle$ is the average of $H$ evaluated at proper time $t$. The volume weighting may also be incorporated into the Langevin approach by taking the appropriate averages as in Ref.~\cite{Gratton,Martin}. Here, an implicit assumption has been made that proper time frame corresponding to different Hubble patches are equivalent. The resulting equation is nonlocal in time, and thus loses many of its nice properties. In particular, there is no guarantee that the entropy ${\cal S}$ always increases. However, one can get around this by working with the non-normalized distribution $\hat{\rho}$ which satisfies the local equation
\be
\frac{\partial \hat{\rho}}{\partial t}=3H\hat{\rho}-\partial_A \hat{\cal J}^A_t,
\ee
with $\hat{\cal J}^A_t$ expressed in terms of $\hat{\rho}$, and then calculate at the end
\be
\rho= \frac{\hat{\rho}}{\int d\phi_A \hat{\rho}}.
\ee
Unlike $\rho$, $\hat{\rho}$ has obvious irreversibility properties, but these are manifest with respect to a different inner product. Working in the Hamilton-Jacobi version, for example, this equation is easiest to interpret by transforming it into a Schrodinger like equation by defining $\hat{\rho}=\exp\({-\int^{\phi} d\phi \frac{8\pi^2M_P^2c_sH_{,\phi}}{H^3}}\) \chi$, with $\chi$ satisfying (with $\beta=1/2$ ordering)
\be
\frac{\partial \chi}{\partial t}=-\hat{K}\chi=- \left [ V_{eff}  -H^{3/2}\frac{\partial}{\partial \phi} \(\frac{H^{3/2}}{8\pi^2} \frac{\partial }{\partial \phi} \) \right ] \chi,
\ee
where
\be
V_{eff}=-3H+\frac{16\pi^2M_P^4c_s^2H_{,\phi}^2}{H^3}+\frac{1}{2}\frac{\partial}{\partial \phi} \( 2M_P^2 c_sH_{,\phi}\).
\ee
It is apparent that the volume weighting term adds a negative contribution to the potential. This comes to dominate precisely when the condition $c_sdS_{\rm ent}/d\lambda <1$ is satsified (at least compared to the second term in $V_{eff}$), i.e. at the same time that quantum corrections kick in. The equilibrium distribution for $\rho$ will be the solution for $\chi$ that falls of as $e^{-\kappa t}$, with the smallest value of $\kappa$ dominating. In other words, this will be the minimum eigenvalue of the operator $\hat{K}$. Intuitively, this equilibrium probability distribution will be peaked near the minimum value of $V_{eff}$, and so correspondingly will be set by the maximum of $H$. In other words, the equilibrium distribution will be peaked at large values of the field $\phi$ precisely at the cutoff of the effective field theory. For the case of k-inflation, a similar conclusion was found in Ref.~\cite{Helmer:2006tz}.

\subsubsection{Phase space Measures}
\label{GT}

Ref.~\cite{GT} proposed the use of a pure phase space measure
on the inflationary system -- which in their analysis consists of a scalar field coupled to gravity in a mini-superspace approximation (for a related phase space discussion in the context of brane inflation see Ref.~\cite{Brandenberger:2003py}). The phase space considered is the true phase space $(\phi,p_{\phi})$ defined on the constrained surface ${\cal H}=0$ by the symplectic two-form
\be
\Omega = d\phi \wedge dp_{\phi}+d\lambda \wedge d\pi_{\lambda}=d\phi \wedge d\( e^{3\lambda }\pi_{\phi}\) - d\lambda \wedge d \(e^{3\lambda} H_{\rm eq}\)
\ee
which is conserved by the equations of motion. This is to be contrasted with the equilibrium phase space we have used in this paper in which the volume factors are taken out
\be
\Omega_{\rm eq}=d\phi \wedge d \pi_{\phi} - d\lambda \wedge d  H_{\rm eq},
\ee
which is not conserved by the equations of motion, hence the presence of the damping terms. However, in the eternally inflating regime, when an equilibrium sets in in which the energy lost through damping is input through the noise, it is $\Omega_{\rm eq}$ that is essentially conserved.

The authors of Ref.~\cite{GT} assume a constant probability measure on this phase space which, in the context of the present language, amounts to assuming $\rho={\rm constant} \times e^{3\lambda}$ and neglecting the diffusion terms in the Fokker-Planck equation.
The phase space is infinite, in general, and so the authors impose a cut off for low spatial curvature, arguing that
universes with very small spatial curvature cannot be observationally distinguished.
Having constructed a Hamiltonian formalism similar to the one we have outlined,
they propose to measure the phase space by analysis of the flux density of classical scalar field trajectories
passing through a particular hypersurface -- the surface of constant $H$.
They are then able to calculate what
fraction of those slowly rolling classical trajectories which pass through this hypersurface near the end of inflation were slow-rolling
throughout the $N$ e-foldings of inflation  that proceeded their passage through this hypersurface. Their conclusion is that the fraction
of always slowly rolling trajectories is $\propto \exp{(-3N)}$. That is, assuming a flat probability of a trajectory beginning at any point in
phase space, the fraction of trajectories that will exhibit canonical slow-roll inflation is suppressed by a large exponential factor.

The volume measure on phase space is given schematically by (though strictly speaking, one should carefully define the surface of integration as in \cite{GT})
\be
\int \int  d\phi d\(\pi_{\phi}e^{3\lambda}\) \sim  \int d\phi \, 2M_P^2|H_{,\phi}|e^{3\lambda} \sim \int dH \, 2M_P^2 e^{3\lambda} .
\ee
If we have $\delta H=\delta H_S$ on some late time surface, then $N$ e-folds earlier we had $\delta H=e^{3N}\delta H_S$. Thus, if we choose to define our measure surface at the end of inflation as in Ref.~\cite{GT}, then their argument follows that the probability of $N$ e-folds of inflation is suppressed by $e^{-3N}$. Whilst the validity of this argument depends on the assumptions that go into it, we see that it conclusions generalize essentially unchanged to completely general models, e.g. those with non-trivial kinetic terms.

\section{Numerical Simulation of Stochastic DBI}\label{numerics}

A complete study of the stochastic behavior of DBI inflation would require us to solve the
Fokker-Planck equation for a variety of parameter choices. However, we can gain
considerable insight into the kinds of phenomenology that the addition of stochastic
dynamics permits by looking at individual field trajectories for particular parameter
choices.

Before describing these results, a brief note on the numerical methods we use.
We solve the Langevin equations for the DBI system using a standard fourth order
Runge-Kutta integrator. We solve the system both with and without the noise
term included. The code adapts the Runge-Kutta step size to keep the evolution of the
classical (no noise) solution well behaved. We refer to Ref.~\cite{langevinnumerics} for a discussion on solving Langevin systems.
Since we will be dealing with Planckian
field values, we solve the equations using variables measured in units of $M_{P}$
(to simplify the formulae, we write the dimensionless Hamiltonian,
${\tilde H}_{\rm eq} =  (H_{\rm eq}/ \sqrt{12}M_{P}^3)$). Throughout the
discussion of our numerical methods, tildes will denote dimensionless quantities.

To treat the noise, we will us a one-noise formulation of the equations. For
this numerical work, the one-noise formulation is superior because, by allowing noise
only in the field's momentum direction, it implicitly enforces the DBI ``speed limit", $|d\phi/dt| \le1/\sqrt{f}$. Allowing
fluctuations in the field value itself leads to many apparent temporary violations of the
``speed limit," making those results harder to interpret. We have done the calculations
in both one- and two-noise cases, nonetheless, and find qualitative agreement between trajectories
derived using both approaches in the
same regions of parameter space.  We expect that these results would harmonize exactly upon
averaging, as would implicitly be done if we were to solve the Fokker-Planck equation
directly. For our single-noise treatment we normalize
the diffusive force term such that
\be
\langle \tilde{\eta}_{\pi_{\phi}} \tilde{\eta}_{\pi_{\phi}} \rangle = 2\tilde{D}^{(1)}_{\pi_{\phi} \pi_{\phi}}  = \frac{1}{4 \pi^2 c_s^2} \tilde{H}_{\rm eq}^4.
\ee
Recall that,
in the DBI framework,
$$
c_s = \frac{1}{\gamma} = \frac{1}{\sqrt{1+f(\phi)\pi_{\phi}^2}},
$$
where $f(\phi)$ is the inverse tension of the D-brane, given by a function of the form
$$
f(\phi) = \frac{\Lambda}{(\phi^2 + b^2)^2},
$$
where $\Lambda$ and $b$ are parameters which can be tuned within the model to
match with data. Typically $\Lambda$ is quite large, and in brane inflation set-ups
is proportional to the number of fluxes present in the Calabi-Yau background. The
so-called mass gap parameter, $b$, is related to physics towards the bottom of the
inflationary throat. For more details on the string theory behind
these parameters, see e.g. \cite{Bean:2007hc}. The standard DBI potential is
given by
$$
V(\phi) = \half m^2 \phi^2 + V_o \left ( 1  - \frac{V_o}{4\pi v} \frac{1}{\phi^{4}} \right ).
$$
For our numerical calculations,
we use a standard Gaussian deviates generator with a variance given by $D_{\pi_{\phi} \pi_{\phi}}$
to draw the stochastic force once per time step.
We assume the noise force is constant during this time step and apply it equally to each of
the Runge-Kutta sub-steps.

\begin{figure}[here] %  figure placement: here, top, bottom, or page
   \centering
   \includegraphics[width=4.5 in]{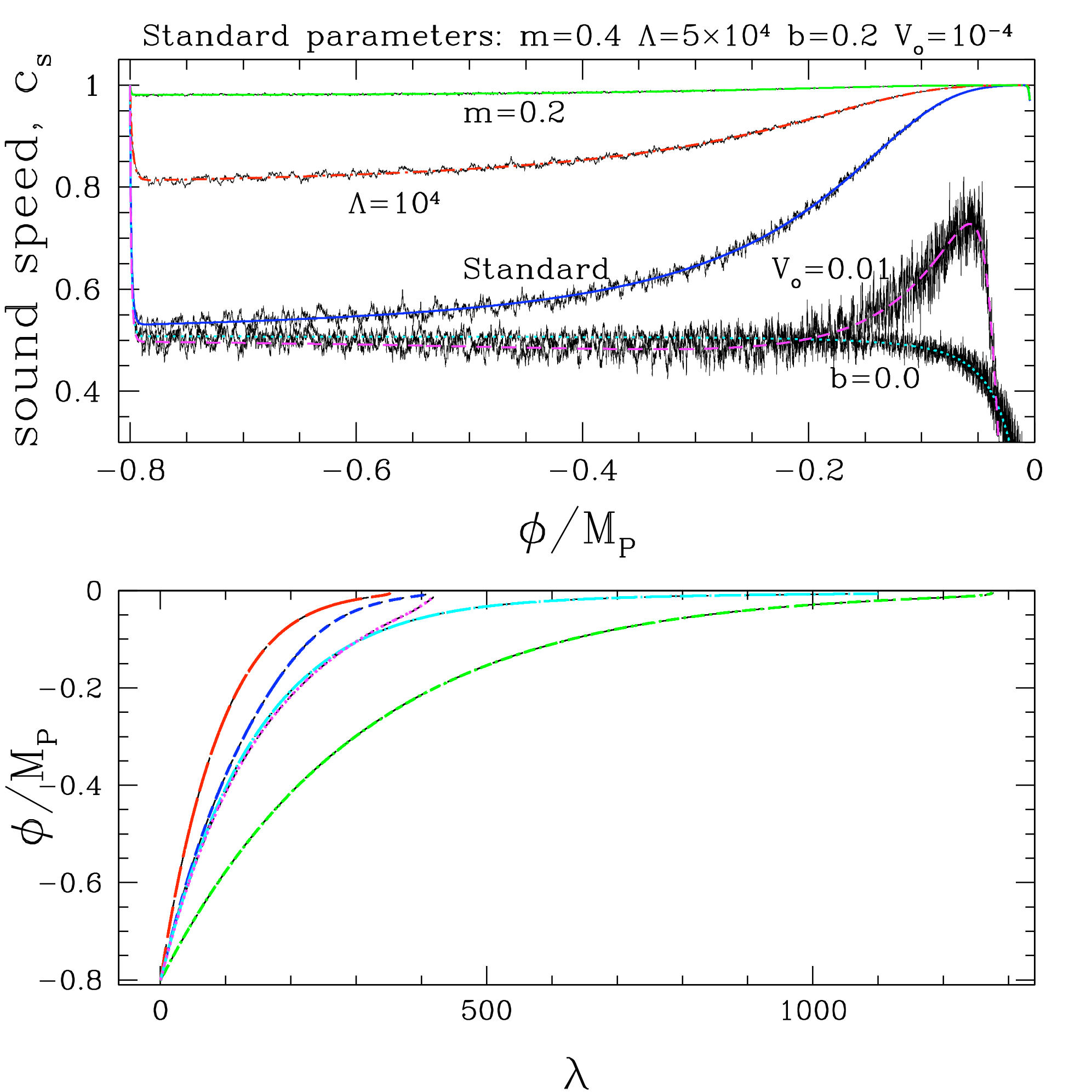}
   \caption{Our standard parameter set is $m=0.4 M_P\, \Lambda=5\times 10^4,$ $b=0.2 M_{P}$,
   $V_o = 10^{-4} M_P^4$, and $v = 10^5$ .  {\bf Upper panel:} the evolution
   of the sound speed over the range between $\phi=-0.8M_{P}$ and $0$. The classical (noise free)
   results are overplotted in colors, while the stochastic results are plotted behind the classical trajectories
    in black. Note that
   the stochastic results have been artificially thinned for readability; not every time step is plotted. In all
   these cases, the stochastic solutions tracked the classical solutions. The dark blue (solid) line
   is the standard parameter set. Each of the other lines has only one changed parameter.
   green (long dash, dot): $m=0.2 M_{P}$; red (long dash, short dash): $\Lambda = 10^4$;  cyan (dotted): $b=0.0$.
   Magenta (dashed): $V_o = 0.01 M_P^4$ {\bf Lower panel:}
   the location of the scalar as a function of e-folding time. The colors correspond to the same
   parameter combinations as in the upper panel.}
  \label{subplanck}
\end{figure}

We present here two parameter regions of interest. In the first, we do not attempt to find
phenomenologically viable inflationary solutions, but merely wish to explore behavior available
to a scalar undergoing stochastic evolution in an effective field theory
described by the DBI action. In this spirit, we decided to see what sorts
of behavior we could obtain using only sub-Planckian values for all parameters and for the
field itself. We selected the parameters by noticing that strongly stochastic behavior that
is also highly relativistic requires, essentially,
\be
\tilde{H}_{\rm eq}^2 \gtrsim \sqrt{\tilde{f}^{-1}(\tilde{\phi})}, \quad
\tilde{\pi}_{\phi} \gtrsim  \sqrt{\tilde{f}^{-1}(\tilde{\phi})},
\ee
where the first inequality ensures strong diffusion, while the second guarantees
$c_s < 1$. We show some typical results for these sorts of parameters in Fig. \ref{subplanck}.

\begin{figure}[here] %  figure placement: here, top, bottom, or page
   \centering
   \includegraphics[width=4.5 in]{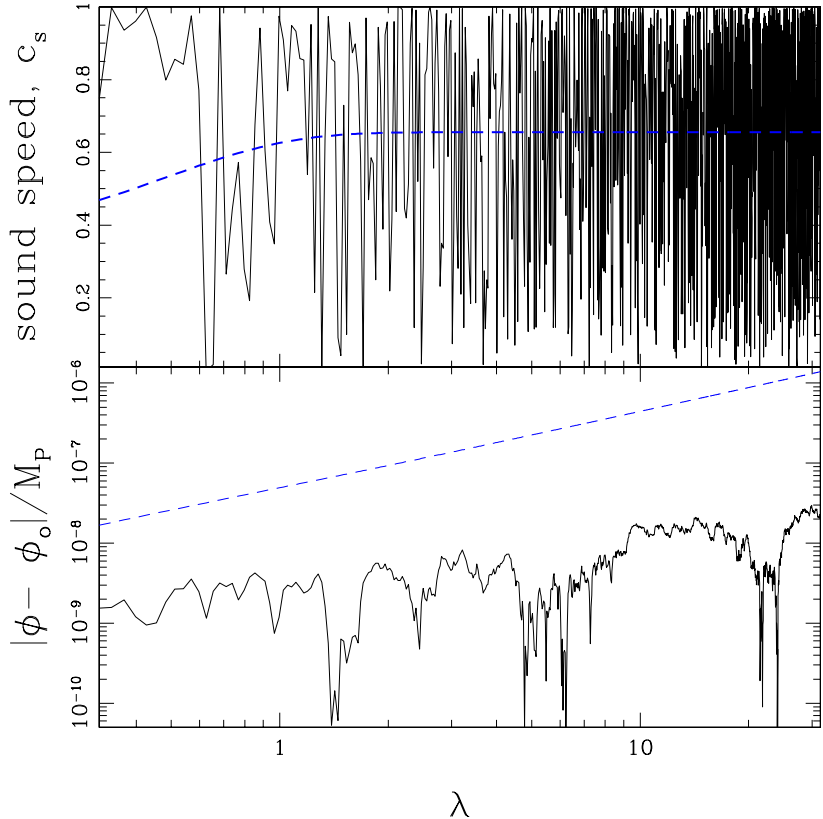}
   \caption{The parameters for this plot are $m=2.4\times10^{-8} M_{P}$, $\Lambda = 4\times10^{23}$,
   $V_o = 2\times10^{-8} M_{P}^4$, $\mu = 0.01$, $ v = 10^{-7} M_{P}^4 $, $b=2.0$. The initial value
   of the $\phi$ field is $\phi_o = 10 M_{P}$. The classical solution (without noise) is plotted
   as a blue dashed line. The stochastic solution is the black solid line, thinned as in Fig. 1.
   The logarithmic scale is used pedagogically, to expand the early time period so that the nature
   of the stochastic trajectory is more readily perceived by the eye.
   {\bf Upper panel:} the speed of sound as a function of efolding time. The frequent spikes to values
   near one represent instances of the stochastic force bringing the field's momentum to values
   near or through zero. {\bf Lower panel:} the evolution of the field away
   from its starting point as a function
   of efolding time. Notice the random walk behavior of the stochastic trajectory.}
   \label{phenom}
\end{figure}

We chose our second set of parameters by comparison with Ref.~\cite{Bean:2007hc}, a
phenomenological study of what parameter ranges for brane inflationary DBI match
the data from WMAP. Most of the parameter sets described there had some interesting
regions of behavior -- we were able to see standard chaotic inflation set in at the appropriate
field values, for instance. But since we are more interested in finding instructive new
stochastic phenomena
than in working out WMAP phenomenology, we chose to plot a related parameter set that is not among
those described in \cite{Bean:2007hc}, but which does reproduce the proper
level of density perturbations for small field values. To achieve this, we have to take
super-Planckian initial conditions for our field values. For
large field values, we expect the first few terms in the DBI potential to be given by
\be
V(\phi) = \half m^2 \phi^2 + V_o \left ( 1 + \mu \phi^4 - \frac{V_o}{4\pi v} \frac{1}{\phi^{4}} \right ),
\ee
where the first term is the usual mass term, the $\phi^{-4}$ is a Coulombic attraction
term arising in brane inflation, and the $\phi^4$ term is a first higher-order correction to the potential,
which we expect when dealing with such large field values.
Our particular parameter choices are:
\begin{eqnarray*}
m= 2.4\times10^{-8} M_{P} \quad & \Lambda = 4\times10^{23} & \quad V_o = 2\times10^{-8} M_{P}^4 \\
\mu = 0.01 \quad & v = 10^{-7} M_{P}^4 \quad & b = 2.0
\end{eqnarray*}
which gives $P(k) \sim 7 \times 10^{-4} \Lambda \, (m^4/M_{P}^4) \sim10^{-10}$ for the part of the
potential dominated by the leading term. In the small $\phi$ region dominated by the $m^2 \phi^2$ potential we easily get as many as $\sqrt{\Lambda/6} \frac{m}{2M_P} \sim 10^4$ e-folds of inflation. For
 $\phi \sim 10 M_{P}$ the supergravity approximation is valid if the string scale,
 $m_{\rm string} \gtrsim 0.04 M_{P}$.
 It is in the regime $\phi \sim 10 M_P$ where we find strong stochastic effects. Some results
are shown in Fig. \ref{phenom}.

The surprise of the results represented in Fig. \ref{phenom} is that a randomly walking, apparently
metastable, stochastic evolution is occurring where, classically, the field is undergoing relativistic,
characteristically DBI evolution. By contrast, in the previous example (Fig. \ref{subplanck}), 
the stochastic trajectories remained quite close to their classical counterparts. 
For this to occur, the value of the tension (given by $1/f(\phi)$)
must be very low, so that the DBI ``speed limit" is saturated even for numerically tiny momenta.
When this occurs, though, the field's speed has such a small upper bound that it can change
signs frequently under stochastic kicks, hence maintaining its stationary position. Interestingly, despite
the frequent passes through zero momentum (and hence a sound speed of unity),
the mean sound speed along the stochastic path is $c_s \sim 0.63$, very close to the
sound speed along the classical trajectory. The existence of such trajectories is hard to interpret
rigorously. Qualitatively, what they teach us is that the conspiracy of noise domination 
and a very small local speed limit can effectively eliminate the possibility of classical evolution
for a DBI scalar field in regions of field-space where these conditions exist. The extreme 
nature of the parameters necessary to achieve this behavior, however, may simply 
indicate that such trajectories are unphysical. It would be interesting, though, nonetheless
to attempt to understand them better, since if they could be realized they would 
constitute a kind of eternal inflation from which there could be no escape.

\section{Stochastic/Thermally-Activated Tunneling}
\label{tunneling}

There are two types of tunneling processes that can occur in nature: thermally-activated tunneling, and quantum tunneling. Thermally-activated tunneling is a classical process
whereby small scale noise fluctuations kick the field over the potential barrier so that it can roll down classically to the true
minimum. In quantum tunneling, by contrast, the field emerges immediately in the new vacuum after following an instanton trajectory
in Euclidean time. These processes coexist at finite temperature, with the system's energy determining which is
dominant, and indeed both can be incorporated into the instanton framework. The stochastic description developed here is
ideally suited to describe thermally-activated tunneling, giving an alternative justification for the Hawking-Moss
instanton. By contrast, the standard stochastic formalism as currently described cannot describe quantum tunneling; that is,
transitions described by Coleman-de Luccia instantons \cite{CdL}. For a broad based discussion
of approaches to tunneling, see Ref. \cite{Linde:1991sk}, and for recent discussions of the relationship between the different tunneling processes see Refs.~\cite{Batra:2006rz, Brown:2007sd}.

Thermally-activated tunneling occurs via diffusion over a barrier (see Kramers \cite{Kramers} for the pioneering work on this). Let us imagine we have a metastable minimum at $\phi_i$ separated from a stable minimum at $\phi_f$ via a potential barrier which is maximized at $\phi_t$.
 The most straightforward way to calculate the tunneling rate is to use the path integral representation of the solution of the Fokker-Planck equation described in Appendix B. In particular, working with the Hamilton-Jacobi version, the probability (density) to go from $\phi_i$ to $\phi_f$ in e-folding time $\lambda_f$ is
\ba
\label{path}
P(\phi_i \rightarrow \phi_f; \lambda_f) & = & \int_{\phi=\phi_i}^{\phi=\phi_f} D \phi \; \exp \left \{ -\int_0^{\lambda_f} d \lambda \frac{2\pi^2}{H^2}\( \frac{d\phi}{d\lambda}+2M_P^2\frac{c_sH_{,\phi}}{H}\)^2 \right \}  \\
&= &  \exp{\( -8\pi^2 \int_{\phi_i}^{\phi_f} d\phi M_P^2\frac{c_sH_{,\phi}}{H^3}   \)} \times \\
&& \int_{\phi=\phi_i}^{\phi=\phi_f} D \phi \; \exp \left \{ -\int_0^{\lambda_f} d \lambda \frac{2\pi^2}{H^2}\(\( \frac{d\phi}{d\lambda}\)^2+\(2M_P^2\frac{c_sH_{,\phi}}{H}\)^2 \)\right \} \nonumber.
\ea
As usual we can perform a saddle-point approximation, where the classical trajectories are solutions to
\be
E= \frac{2\pi^2}{H^2}\(\( \frac{d\phi}{d\lambda}\)^2-\(2M_P^2\frac{c_sH_{,\phi}}{H}\)^2 \),
\ee
where $E$ is a conserved `energy' variable. In general the solutions to this equation are unstable due to the negative potential, but the unstable one have much larger `action' (where by action we means the expression in the path integral (\ref{path})) and are consequently much less probable. The unstable trajectories are those allowed by thermal kicks and allow the field to roll up the hill, whereas the usual trajectories that satisfy the classical equations of motion contribute to the probability with unit measure.

It is easy to demonstrate that the minimum action trajectory corresponds to $E=0$, which means that the trajectories that go from $\phi_i$ to $\phi_t$ evolve according to the e-folding time reversed equations of motion
\be
\frac{d\phi}{d\lambda}=+2M_P^2\frac{c_sH_{,\phi}}{H}.
\ee
This means that these trajectories are repellers by the argument of Sec.~\ref{Hamilton-Jacobi}, i.e. $\delta H \sim e^{3\lambda}$.
Ignoring fluctuation determinant factors, the tunneling rate (measured per e-folding time) is given approximately by
\be
\Gamma=\frac{1}{\int_{\phi_i}^{\phi_f} d\phi |2M_P^2\frac{c_sH_{,\phi}}{H}|^{-1}} \exp\( -8\pi^2 \int_{\phi_i}^{\phi_f} d\phi M_P^2\( \frac{c_sH_{,\phi}}{H^3}+ \left |\frac{c_sH_{,\phi}}{H^3} \right | \)\)
\ee
Thus the exponential suppression comes entirely from the part of the path with $H_{,\phi} \ge 0$. This
path is classically forbidden in the absence of fluctuations, so in this region we find
\be
\Gamma=\frac{1}{\int_{\phi_i}^{\phi_f} d\phi |2M_P^2\frac{c_sH_{,\phi}}{H}|^{-1}} \exp\( -16\pi^2 \int_{\phi_i}^{\phi_t} d\phi M_P^2\( \frac{c_sH_{,\phi}}{H^3} \)\).
\ee
For $c_s=1$ this is exactly the standard Hawking-Moss result $\Gamma \sim \exp\( S_{\rm ent}(\phi_t)-S_{\rm ent}(\phi_i)\)$. It is clear that if $c_s \ll 1$ in the region between $\phi_i$ and $\phi_t$ the argument
of the exponential is significantly enhanced. In particular, in the ultra-relativistic limit $c_s \sim 1/(2M_P^2f^{1/2} |H_{,\phi}|)$ and so
\be
\Gamma \sim \exp\(-8\pi^2 \int_{\phi_i}^{\phi_t} d\phi \sqrt{\frac{27M_P^6}{fV^3}} \).
\ee
As we increase the number of fluxes $\Lambda$, thus sending the tension $T =1/f$ towards zero, the exponential suppression tends to unity, implying that thermal tunneling can be significantly enhanced in the relativistic regime. A similar effect was found in Ref.~\cite{CdL} with regards to quantum tunneling,
albeit through a very different mechanism and with a very different quantitative form.

There is a second way to get this result which follows more closely the original arguments in \cite{Kramers}. The key point is that if the tunneling rate is low, then there is usually sufficient time for an approximate equilibrium to set in for the field in the metastable well. Thus the probability density will be given approximately by the equilibrium configuration found earlier. The rate is then given by $\Gamma={{\cal J}_{\phi}}/{\rho_{\rm eq}(\phi_i)}$.
A simple calculation along the lines of Sec.~5.10.1 in Ref.~\cite{Risken:1989fp} shows that the flux ${\cal J}_\phi$ is dominated by $\rho_{\rm eq}(\phi_t)$, and so we have
\be
\Gamma \approx \frac{\rho_{\rm eq}(\phi_t)}{\rho_{\rm eq}(\phi_i)},
\ee
which is the same result.

It is crucial to stress that this formula for the tunneling rate depends on the variable $J$ implicitly though the Hamilton-Jacobi function $H(\phi,J)$. Thus the tunneling rate retains a memory of the initial state which has not been washed out by stochastic fluctuations. Since the trajectories that go up the potential barrier are repellers, this can have a significant effect on the tunneling rate.

Note that the above path integral derivation can easily be extended to tunneling over multiple barriers with the total probability being suppressed by the product of the exponential suppressions for tunneling to the bottom and top of each barrier.

\section{Summary}

Recent progress toward building phenomenologically viable inflationary models in string theory
has many consequences, chief among them the need to grapple with the landscape of vacua.
Together with the rise in popularity of other non-standard inflationary models,
the time is ripe to reconsider some of the standard techniques used in exploring the full implications of inflation.
In this article we review some the salient features of the stochastic inflationary framework, taking a very general,
model independent approach. Using our formalism, we specialize to non-minimal kinetic models
such as DBI inflation that appear naturally in a stringy framework. We use this as an example
of the concrete uses of our formalism, and were able to discover novel stochastic behavior for
DBI inflation.  By solving the stochastic dynamics numerically, we have found that metastable
equilibrium states can exist in which the field is trapped by quantum corrections at some fixed value.
We also find that the rate for Hawking-Moss type thermal tunneling can be significantly enhanced in regions of
phase space where relativistic DBI effects are important.

\acknowledgments
The work of A.J.T. and M.W.  was supported by the Perimeter Institute for Theoretical
Physics.  Research at the Perimeter Institute is supported by the Government
of Canada through Industry Canada and by the Province of Ontario through
the Ministry of Research \& Innovation.
%We thank the Abdus Salam ICTP and the University of Sussex for
%hospitality during the completion of this work.
We thank N. Afshordi, C. de Rham, G. Geshnizjani, S. Sarangi, S. Shandera, H. Tye, N. Turok and I. Wasserman for helpful discussions, and illy coffee for stimulation.

\appendix

\section{Hubble Parameter as Equilibrium Hamiltonian}

Consider an arbitrary system of matter fields $\phi_{\alpha}$ coupled to Einstein-Hilbert gravity
\be
S=\int d^4x \sqrt{-g} \(\frac{1}{2}M_{P}^2 R+{\mathcal L}_M(\phi_{\alpha},\partial_{\mu}\phi_{\alpha},g_{\mu\nu}) \).
\ee
In the mini-superspace approximation this reduces to (for fixed comoving volume)
\be
S=\int dt L = \int N dt e^{3\lambda} \(-3M_{P}^2 \frac{\dot{\lambda}^2}{N^2}+{\mathcal L}_M \(\phi_{\alpha},\frac{1}{N}\frac{\partial \phi_{\alpha}}{\partial t}\)\).
\ee
Defining the canonical conjugate with the inverse volume factor
\be
\pi_{\lambda}=e^{-3\lambda}\frac{\partial \mathcal{L}}{\partial {(\dot{\lambda}/N)}}=-6M_P^2\frac{\dot{\lambda}}{N}=-6M_{P}^2H, \quad
\pi_{\phi_{\alpha}}=e^{-3\lambda} \frac{\partial \mathcal{L}}{\partial ( \dot{\phi}_{\alpha}/N)},
\ee
so that the action can be re-expressed as
\be
S=\int dt e^{3\lambda} \(\pi_{\lambda}\dot{\lambda}+\pi_{\phi_{\alpha}}\dot{\phi}_{\alpha}-N{\cal H}(\phi_{\alpha},\pi_{\phi_{\alpha}},\pi_{\lambda}) \).
\ee
Solving the Hamiltonian constraint ${\cal H}=0$ for $\pi_{\lambda}$ and substituting back in gives the reduced phase space action expressed in e-folding time
\be
S=\int d\lambda \, e^{3\lambda} \( {\pi}_{\phi_{\alpha}}\frac{d\phi_{\alpha}}{d\lambda}+{\pi}_{\lambda}\),
\ee
where the momenta are now implicitly evaluated on the constraint surface.
This is exactly of the general scaling form where we identify $H_{\rm eq}=-{\pi}_{\lambda}$ as the equilibrium Hamiltonian (generator of $\lambda$ translations). Thus in general $H_{\rm eq}=6M_{P}^2H$. As apparent from Sec.~\ref{Hamilton-Jacobi}, this result is closely related to the fact that the Hamilton-Jacobi function is directly proportional to the Hubble parameter.

\section{General Relation between Langevin and Fokker-Planck Equations}

Consider a generalised phase space damped Langevin system described by the equations
\be
\label{generallangevin}
\frac{d\phi_A}{d\lambda}=\Omega_{AB}\frac{\partial H_{\rm eq}}{\partial \phi_A}-\gamma_{A}{}^{B}\phi_B+\eta_A
\ee
where $\Omega_{AB}$ is the phase space symplectic form,  $\gamma_A{}^B$ is a generalization of the damping parameter,
and $\eta_A$ is the stochastic noise. If this noise is Markovian and Gaussian, we may always rewrite this equation in the form of a Fokker-Planck equation. This is facilitated by means of path integrals (see \cite{Kleinert} for an excellent review of this). Since the noise is Gaussian with correlators
\be
\langle \eta_A(\lambda) \eta_B(\lambda') \rangle = 2 D_{AB}\delta(\lambda-\lambda'),
\ee
then the average of a physical quantity may be obtained from the path integral
\be
\langle {\mathcal O} \rangle = \int D\eta_A {\mathcal O}(\eta_A) \exp\({-\frac{1}{4}\int d\lambda \eta_AK^{AB}\eta_B}\)
\ee
where $K^{AB}$ is the inverse of $D_{AB}$ which we shall (temporarily) assume to be invertible. We can view equations \ref{generallangevin} as defining the noise in terms of $\phi_A$ and so change variables in the path integral to
\be
\langle {\mathcal O} \rangle = \int D\phi_A {\cal M} \, {\mathcal O}(\phi_A) e^{{-\frac{1}{4}\int d\lambda \(\frac{d\phi_A}{d\lambda}-\Omega_{AC}\frac{\partial H_{\rm eq}}{\partial \phi_C}+\gamma_{A}{}^{C}\phi_C\)K^{AB}\(\frac{d\phi_B}{d\lambda}-\Omega_{BD}\frac{\partial H_{\rm eq}}{\partial \phi_D}+\gamma_{B}{}^{D}\phi_D\)}},
\ee
where ${\cal M}$ is the change of measure determinant which can be dealt with using the usual Fadeev-Popov ghosts \cite{Kleinert}. This is a subleading effect that we can safely ignore in the present argument. This is closely connected with the ordering issue of the derivatives in the Fokker-Planck equation.
At this point we can introduce auxiliary conjugate momenta $P^A$ (N.B. no relation to the $\pi_{\alpha}$) and write
\begin{eqnarray}
\label{supercanonical}
\langle {\mathcal O} \rangle &=& \int D\phi_A D P^A \, {\mathcal O}(\phi_A) \exp\({i\int d\lambda P^A\(\frac{d\phi_A}{d\lambda}-\Omega_{AB}\frac{\partial H_{\rm eq}}{\partial \phi_B}+\gamma_{A}{}^{B}\phi_B\)-\int d\lambda P^A D_{AB} P^B }\), \nonumber \\
&=& \int D\phi_A D P^A \, {\mathcal O}(\phi_A) \exp\({i\int d\lambda \( P^A\frac{d\phi_A}{d\lambda}-\hat{H}(\phi_A,P^A) \)}\),
\end{eqnarray}
where the enlarged phase space Hamiltonian is
\be
\hat{H}(\phi_A,P^A)= P^A \(\Omega_{AB}\frac{\partial H_{\rm eq}}{\partial \phi_B}-\gamma_{A}{}^{B}\phi_B\)-i P^A D_{AB} P^B .
\ee
The Fokker-Planck equation is simply the Schrodinger equation associated with this path integral, obtained by the usual replacement $P_A \rightarrow -i\frac{\partial}{\partial \phi_A}$ (modulo operator ordering issues),
\be
\frac{\partial \rho}{\partial \lambda}=-\frac{\partial}{\partial \phi_A} \( \(\Omega_{AB}\frac{\partial H_{\rm eq}}{\partial \phi_B}-\gamma_{A}{}^{B}\phi_B\)\rho -D_{AB}\frac{\partial}{\partial \phi_B}\rho \).
\ee
In the case where $D_{AB}$ is not invertible this result is still valid, and all that happens is that the Langevin equations that do not directly have noise sources give rise to delta functions which enforce the classical equations of motion in equation \ref{supercanonical}. The above path integral representations far from being  just a useful mathematical trick can be derived from first principles by coarse-graining the {\it in-in}/Schwinger-Keldysh path integral along the lines considered in the stochastic gravity approach \cite{Hu}; see also \cite{Goncharov:1986ua}.

\end{document}